# Numerical solution of the exact background collisional Boltzmann equation for dark matter-baryon scattering

Suroor Seher Gandhi [*] and Yacine Ali-Haïmoud[†]
*Center for Cosmology and Particle Physics, New York University, 726 Broadway, New York, NY 10003, USA*
(Dated: September 1, 2022)

Linear cosmological observables can be used to probe elastic scattering of dark matter (DM) with baryons. Availability of high-precision data requires a critical reassessment of any assumptions that may impact the accuracy of constraints. The standard formalism for constraining DM-baryon scattering pre-recombination is based on assuming a Maxwell-Boltzmann (MB) velocity distribution for DM. This assumption is not always physically justified, and does not allow for probing DM self-interactions simultaneously with its interactions with baryons. Lifting the MB assumption requires solving the full collisional Boltzmann equation (CBE), which is highly non-trivial. Earlier work proposed a more tractable Fokker-Planck (FP) approximation to the CBE, but its accuracy remained unknown. In this paper, we numerically solve the exact CBE for the first time, in a homogeneous expanding background. We consider DM-baryon scattering cross-sections that are positive power-laws of relative velocity. We derive analytical expressions for the collision operator in the case of isotropic differential scattering cross-sections. We then solve the background CBE numerically, and use our solution for the DM velocity distribution to compute the DM-baryon heat-exchange rate, which we compare against those obtained with the MB assumption and FP approximation[a]. Over a broad range of DM-to-baryon mass ratios, we find that the FP approximation leads to a maximum error of 17%, significantly better than the up to 160% error introduced by the MB assumption. While our results strictly apply only to the background evolution, the accuracy of the FP approximation is likely to carry over to perturbations. This motivates its implementation into cosmological Boltzmann codes, where it can supersede the much less accurate MB assumption, and allow for a more general exploration of DM interactions with baryons and with itself.

## I. INTRODUCTION

The synchronous advance of cosmology, astrophysics, and particle physics is now being propelled forward like never before. With unprecedented precision, experiments can access increasingly distant realms of the universe— seeking evidence of existing theoretical predictions, while inevitably observing phenomena yet to be explained by theory. The quest to understand dark matter (DM) is one of the key motivations driving this concerted effort.

Interactions between Standard Model (SM) particles and DM could provide insights into the nature of DM. Probes of such interactions include direct-detection experiments searching for nuclear recoil from scattering events [for a review, see 1]. The parameter space probed by these experiments is currently limited to DM masses above a few GeV for DM-nucleon scattering, and above a few MeV for DM-electron scattering [2]; moreover, such experiments are only sensitive to cross-sections below the direct-detection "ceiling" [e.g., 3–5]. A natural extension beyond the terrestrial search for DM-baryon scattering is to analyze its effects on cosmological and astrophysical observables [6].

Heat and momentum exchange between DM and baryons due to their elastic scattering modifies the cosmological thermal history and structure formation away from the standard $\Lambda$CDM paradigm. The cosmic laboratory offers probes of such deviations across the entire history of the universe, constraining parameter space complementary to that of direct detection. DM candidates which scatter with baryons pre-recombination can extract heat from the photon-baryon plasma, thus generate spectral distortions in the Cosmic Microwave Background (CMB) [7, 8]. Additionally, such scattering leads to momentum exchange between the DM and photon-baryon fluids, which can smooth out the growth of small-scale primordial fluctuations over time, alter lensing, polarization, and temperature anisotropies of the CMB, as well as small-scale matter overdensities [e.g., 9–15]. Effects on the small-scale linear matter power spectrum propagate to non-linear, low-reshift observables ($z \lesssim 10$) such as the Ly-$\alpha$ forest and galaxy distribution [9–11, 14–20]. Efficient DM-baryon interactions post-recombination can modify the spin temperature of neutral hydrogen, in turn affecting the 21-cm signal [6, 21–25]. Lastly, in addition to these cosmological probes, several astrophysical tests have been proposed [26–33].

Almost all existing limits on DM-baryon scattering obtained using the astrophysical and cosmological probes mentioned above rely on the key assumption that the DM has a thermal, Maxwell-Boltzmann (MB) velocity distribution[1]. This assumption dramatically simplifies calcu-lations, as it allows one to compute heat and momentum exchange rates for DM-baryon scattering by solving ordinary differential equations. The MB assumption, however, is strictly justified only when the DM particles either self-interact efficiently and are hence thermalized, or efficiently scatter with baryons so as to be equilibrated with them. Typically, DM is initially coupled to baryons and eventually decouples from them. In such cases, a necessary condition for the MB assumption to be accurate is that, at least until some time after DM-baryon decoupling, the DM-DM interaction rate remains much greater than the expansion rate, and thus, much greater than the DM-baryon interaction rate. Such a condition would be satisfied, for instance, if the DM is part of a larger and possibly strongly-interacting dark sector, which is itself weakly coupled to the SM. But this is certainly not a universal prediction of DM models, and it is also possible that DM self-interactions are suppressed relative to DM-baryon interactions (see e.g. Refs. [36, 37] for an explicit model). In such cases, the MB assumption may result in inaccurate predictions for heat and momentum exchange rates, and ultimately for astrophysical and cosmological observables probing DM-baryon interactions.

---

[*] ssg487@nyu.edu
[†] yah2@nyu.edu
[a] The code used for this work is available at the GitHub respository suroorseherg/dm-b_scatt.

[1] One notable exception is Ref. [34], where the authors account for the possibility of a non-thermal background velocity distribution for a DM particle with a small effective electric charge, produced by the freeze-in mechanism [35].

With multiple observational sources of high-precision data at our disposal (especially *Planck*'s CMB power spectra [38]), ensuring the accuracy of methods used to calculate DM-baryon scattering constraints is critical. Beyond improving modeling accuracy, relaxing the MB assumption would open up the possibility of simultaneously and agnostically probing DM-baryon interactions and DM self-interactions, hence exploring DM properties in more detail. This will be especially relevant for upcoming and planned CMB missions, which could probe DM-baryon interactions well below current cosmological limits [39]. In the case of a positive detection of DM-baryon interactions, the natural next step would be to study DM self-interactions, which is not doable within the existing framework for linear-cosmology observables.

Going beyond the MB assumption is a highly non-trivial task, as it requires solving the collisional Boltzmann equation (CBE) for the full DM velocity distribution, rather than merely following its average and variance. Currently, a method that solves the inhomogeneous and anisotropic CBE and can be integrated into Boltzmann codes such as CAMB [40] or CLASS [41] does not exist. A formalism to approximate the DM-baryon collision operator with a Fokker-Planck (FP) diffusion operator was developed in Ref. [42] (hereafter, Paper-I). Although it is significantly more complex than the MB approximation, the numerical solution of the Boltzmann-FP equation is much more tractable than that of the full CBE. In Paper-I it was shown that, in an isotropic and homogeneous background, up to $\mathcal{O}(1)$ differences can arise between the heat-exchange rates calculated using the MB assumption and those obtained from solving the Boltzmann-FP equation. While the FP approximation is likely more accurate than the MB assumption, as it involves solving – even if approximately – for the full DM phase-space density rather than assuming a specific shape, its accuracy remains to be quantified. Indeed, for DM particles lighter than the SM particles with which they scatter, DM-baryon scattering is not a diffusive process, and the FP approximation could be, in principle, just as inaccurate as the much simpler MB assumption. The regime of light DM is where an exact implementation of the CBE would be especially useful.

In this paper, we take the first step towards quantifying the accuracy of the FP approximation against an exact method. We do so by solving the full CBE for an isotropic DM velocity distribution in a spatially homogeneous (unperturbed) background, accounting for DM-baryon elastic scattering through an exact collision operator. We limit ourselves to the regime of negligible DM self-interactions, thereby obtaining the *maximal* error in the FP or MB methods. We consider DM scattering elastically with baryons with a cross-section scaling as a power law of relative velocity, $\sigma(v) \propto v^n$, where $n$ is an integer. This class of models has been widely analyzed and constrained with cosmological data [e.g., 7, 8, 11–15, 21–23, 43]. We focus on models with even and positive powers $n \in \{0, 2, 4\}$ for simplicity. Such models arise from effective field theory and are studied in the context of direct detection [13, and references therein]. In these models, elastic DM-baryon scattering is most efficient early in the radiation-dominated era and becomes inefficient well before recombination ($z \gg z_{\rm rec}$), for currently allowed cross-sections. Models with $n < 0$ exhibit late-time DM-baryon scattering ($z \lesssim z_{\rm rec}$), when bulk DM-baryon relative velocity becomes supersonic and structure formation is no longer linear, leading to additional complications.

For the first time, we derive analytical expressions for differential elastic scattering rates, as a function of initial and final DM velocities, in the case of isotropic differential scattering cross-sections. We express these rates in terms of rescaled velocities, allowing us to factorize the collision operator into a pre-computable time-independent piece, and an overall time-dependent prefactor. We then numerically solve the full CBE for an isotropic DM velocity distribution, in a homogeneous expanding background. From this solution, we extract the DM-baryon heat-exchange rate, and compare it with those obtained within the MB and FP approximations. For the broad range of DM-baryon mass ratios we consider, we find that the FP approximation is systematically and significantly more accurate than the MB approximation: the heat-exchange rate obtained in the FP approximation differs by no more than 17% from the one obtained from the exact solution of the CBE, in contrast with the up to 160% inaccuracy of the MB approximation.

It is important to note that our quantitative results only apply to the *background* evolution and heat-exchange rate, not to the momentum-exchange rate which is most relevant to all structure-based observables. Still, our findings should hold *qualitatively* for momentum-exchange rate, and suggest important impli-cations. First, the at most order-unity error we find for the background heat-exchange rate in the MB approximation is reassuring for existing upper limits relying on this simple method, as there is no reason to expect momentum-exchange rates to be vastly more inaccurate. Second, our results bode well for the higher accuracy of the FP approach in general, and motivate the implementation of the perturbed Boltzmann-FP hierarchy derived in Paper-I in cosmological Boltzmann codes. Third, and most importantly, the non-negligible difference between the strong-self-interaction limit (captured accurately in the MB approximation) and the negligible-self-interaction limit (which only the exact CBE can accurately describe) indicate that there is room for exploring DM self-interactions on top of its interactions with baryons. Beyond a mere improvement of accuracy, the FP or exact methods thus hold the promise of opening new avenues for testing properties of DM which are not accessible within the MB approximation. This will be especially relevant with the high-precision upcoming surveys like the Rubin and Simons Observatories [44, 45], for which it will be imperative to develop a not only more accurate, but also a more general formalism for DM interactions.

The remainder of this paper is organized as follows. In Sec. II, we introduce our notation, the general theoretical formalism, and explain the MB and FP methods. In Sec. III, we describe our method of exactly solving the CBE. We describe and discuss our results in Sec. IV, and conclude in Sec. V. Appendices A and B provide derivations of the analytic differential scattering rates, for isotropic differential scattering cross-sections, and Appendix C provides analytic expressions for the dimensionless rates of velocity diffusion.

## II. THEORY

### A. Basic notation

Throughout the paper, we use '$\chi$' to represent DM, assumed to be non-relativistic, '$s$' for a non-relativistic SM scatterer like an electron, proton or helium nucleus[2], and '$b$' for the combined fluid of the SM scatterer species. We denote the DM and scatterer's number densities by $n_\chi$ and $n_s$, respectively, and their mass densities by $\rho_\chi = m_\chi n_\chi$ and $\rho_s = m_s n_s$, where $m_\chi$ is the DM particle mass, and $m_s$ is the scatterer's mass. We also define $M \equiv m_s + m_\chi$ to be the total mass of the DM-scatterer system. We denote scatterer velocities by $\vec{v}_s$ and DM velocities by $\vec{v} \equiv \vec{v}_\chi$, dropping the $\chi$ subscript for compactness. We label quantities calculated via our exact implementation, the MB assumption, or FP approxima-tion with the superscripts 'ex', 'MB', and 'FP', respectively. General quantities which are not associated with any particular method (exact, MB, or FP) are denoted without any superscript.

### B. Relevant quantities and observables

Cosmological probes of DM-baryon scattering are very sensitive to two quantities: the rates of heat and momentum exchange between DM and baryons. While these are not observables *per se*, they have such a direct impact on various observable quantities that we refer to them, for short, as "observables" – explicitly, the heat-exchange rate is the most closely related to CMB spectral distortions [7, 8], while the momentum-exchange rate is the quantity most relevant to CMB-anisotropy and large-scale structure observables [11, 13].

Both heat- and momentum-exchange rates depend on $f_\chi(\vec{v})$, the probability distribution function of DM velocities, normalized as[3]

$$\int d^3v \; f_\chi(\vec{v}) = 1 \;. \tag{1}$$

Given $f_\chi(\vec{v})$, we may define the DM bulk velocity $\vec{V}_\chi$ and temperature $T_\chi$ as follows

$$\vec{V}_\chi \equiv \int d^3v \; \vec{v} \; f_\chi(\vec{v}), \tag{2}$$

$$T_\chi = \frac{1}{3} m_\chi \int d^3v \; (\vec{v} - \vec{V}_\chi)^2 f_\chi(\vec{v}). \tag{3}$$

The observables of interest are then

$$\dot{\vec{P}}_\chi \big|_{\text{scat}} \equiv \rho_\chi \dot{\vec{V}}_\chi \big|_{\text{scat}}, \tag{4}$$

$$\dot{\mathcal{Q}}_\chi \big|_{\text{scat}} \equiv \frac{3}{2} n_\chi \dot{T}_\chi \big|_{\text{scat}}, \tag{5}$$

which are the volumetric rates of momentum- and heat-exchange, respectively. In these expressions, the subscript "scat" means that one is to only account for the rate of change of $\vec{V}_\chi$ and $T_\chi$ due to DM-baryon scattering (and not, e.g., due to cosmological expansion).

Before we proceed, let us introduce some useful notation. For a particle of mass $m$ and velocity $\vec{v}$, we denote by $f^{\text{MB}}(\vec{v}; \vec{V}, T/m)$ the Maxwell-Boltzmann distribution with bulk velocity $\vec{V}$ and temperature $T$, given by

$$f^{\text{MB}}\left(\vec{v}; \vec{V}, T/m\right) \equiv \frac{1}{(2\pi T/m)^{3/2}} \exp\left(-\frac{m}{2T}(\vec{v} - \vec{V})^2\right). \tag{6}$$

Baryons efficiently scatter with one another and thus have thermal (MB) velocity distributions, all with the same bulk velocity $\vec{V}_e = \vec{V}_p = \vec{V}_{\text{He}} \equiv \vec{V}_b$, and all with the same temperature $T_e = T_p = T_{\text{He}} \equiv T_b$:

$$f_s(\vec{v}_s) = f^{\text{MB}}\left(\vec{v}_s; \vec{V}_b, T_b/m_s\right), \qquad s \in \{e, p, \text{He}\}. \quad (7)$$

---

[2] We restrict ourselves to redshifts $z \ll 10^9$ when the CMB temperature is well below the mass of each of these species.

[3] The DM phase-space density is then $(n_\chi/m_\chi^3) f_\chi(\vec{v})$.

In general DM cannot be assumed to have a MB distribution of velocities. However, in the limit that it is tightly coupled to baryons (e.g. at sufficiently early times), its distribution does tend to the equilibrium MB distribution with bulk velocity $\vec{V}_b$ and temperature $T_b$, which we denote for short by

$$f_\chi^{\text{eq}}(\vec{v}) \equiv f^{\text{MB}}(\vec{v}; \vec{V}_b, T_b/m_\chi). \quad (8)$$

Importantly, momentum- and heat-exchange rates are entirely determined by the *deviations* of $f_\chi$ from equilibrium with the baryons, $\Delta f_\chi \equiv f_\chi - f_\chi^{\text{eq}}$.

While the DM distribution function is not itself observable, its evolution does determine the momentum and heat-exchange rates, as we will see in detail below. In the following sections we describe three different approaches to computing $f_\chi$, and thus observable quantities.

### C. Exact method: collisional Boltzmann equation

#### 1. General equation

The DM velocity distribution evolves according to the collisional Boltzmann equation (CBE),

$$n_\chi^{-1} \frac{d}{dt}\bigg|_{\text{free}} [n_\chi f_\chi(\vec{v})] = C_{\chi s}[f_\chi](\vec{v}) + C_{\chi\chi}[f_\chi](\vec{v}), \quad (9)$$

where $d/dt|_{\text{free}}$ is the time derivative along free (collisionless) trajectories, $C_{\chi s}[f_\chi]$ is the DM-baryon collision operator, and $C_{\chi\chi}[f_\chi]$ is the DM-DM collision operator. In a homogeneous universe expanding with Hubble rate $H$, for non-relativistic particles we have, explicitly,

$$\frac{d}{dt}\bigg|_{\text{free}} \equiv \frac{\partial}{\partial t} - H\vec{v} \cdot \frac{\partial}{\partial \vec{v}}. \quad (10)$$

The collision operators account for the evolution of $f_\chi(\vec{v})$ beyond free-streaming, due to scattering processes. The DM-baryon scattering operator is a linear integral operator, given by

$$C_{\chi s}[f_\chi](\vec{v}) = \int d^3v' \Big[ f_\chi(\vec{v}')\Gamma_{\chi s}(\vec{v}' \to \vec{v}) \\ - f_\chi(\vec{v})\Gamma_{\chi s}(\vec{v} \to \vec{v}') \Big], \quad (11)$$

where $\Gamma_{\chi s}(\vec{v} \to \vec{v}')$ is the differential rate at which DM with initial velocity $\vec{v}$ scatters into final velocity $\vec{v}'$, per final velocity volume. The first term in Eq. (11) represents the rate at which $\chi$-particles acquire velocity $\vec{v}$ after scattering with baryons, and the second term represents the rate at which $\chi$-particles with initial velocity $\vec{v}$ scatter and acquire a different final velocity $\vec{v}' \neq \vec{v}$. This operator conserves probability (or equivalently the number of particles), as can be seen by evaluating its effect on $f_\chi(\vec{v})$ integrated over all $\vec{v}$ at a given instance in time:

$$\int d^3v \, C_{\chi s}[f_\chi](\vec{v}) = 0. \quad (12)$$

Given that baryons are thermalized, the differential scattering rates satisfy *detailed balance*:

$$f_\chi^{\text{eq}}(\vec{v})\Gamma_{\chi s}(\vec{v} \to \vec{v}') = f_\chi^{\text{eq}}(\vec{v}')\Gamma_{\chi s}(\vec{v}' \to \vec{v}). \quad (13)$$

It follows that

$$C_{\chi s}[f_\chi^{\text{eq}}](\vec{v}) = 0, \quad (14)$$

which is a restatement of the fact that if DM is in equilibrium with baryons (i.e., has a MB distribution at temperature $T_\chi = T_b$), its distribution is not changed by scattering with baryons.

We will not explicitly compute the DM-DM scattering operator $C_{\chi\chi}[f_\chi](\vec{v})$ in this paper, but let us simply mention some of its properties. It is an integral operator *quadratic*, and thus non-linear, in $f_\chi$. It also conserves DM number, and vanishes when applied to *any* MB distribution, at an arbitrary temperature $T_\chi$ (which may or may not be the same as $T_b$).

To compute the momentum- and heat-exchange rates, in general, one therefore has to first solve the collisional Boltzmann equation, which is an *integro-differential equation*. This equation is linear in $f_\chi$ when DM-DM interactions are negligible.

#### 2. Background equation

Both baryon and DM distribution functions can be split into a homogeneous and isotropic background piece, and an inhomogeneous, anisotropic perturbation: $f_\chi(\vec{v}, \vec{x}) = \overline{f}_\chi(v) + \delta f_\chi(\vec{v}, \vec{x})$, and similarly for baryons. For small perturbations, as is the case in the early Universe, the evolution of the background distributions does not depend on perturbations (the converse, however, is not true). While the perturbations determine the rate of momentum exchange, most relevant to structure-based observables, their treatment is more complex than that of the background. As a first step, we focus on the evolution of $\overline{f}_\chi(v)$ in this work, from which we will extract the background heat-exchange rate.

We may convert the 3-dimensional Boltzmann equation into a 1-dimensional equation, by defining the following quantities:

$$f_\chi^{\text{1D}}(v) \equiv 4\pi v^2 \, \overline{f}_\chi(v), \quad (15)$$

$$\Gamma_{\chi s}^{\text{1D}}(v \to v') \equiv (v')^2 \int \frac{d^2\hat{v}}{4\pi} \int d^2\hat{v}' \, \Gamma_{\chi s}(\vec{v} \to \vec{v}'). \quad (16)$$

The 1-dimensional velocity distribution $f_\chi^{\text{1D}}(v)$ is such that $\int dv \, f_\chi^{\text{1D}}(v) = 1$, and the differential scattering rate $\Gamma^{1D}_{\chi s}(v \to v')$ has dimensions of rate per final velocity magnitude interval. Note that, given that the background baryon distribution is isotropic, $\Gamma_{\chi s}(\vec{v} \to \vec{v}')$ is a function of $v, v'$ and $\hat{v} \cdot \hat{v}'$ alone, and as a consequence it suffices to integrate $\Gamma_{\chi s}(\vec{v} \to \vec{v}')$ over the directions of final velocities to obtain $\Gamma^{1D}_{\chi s}(v \to v')$, i.e. the averaging over $\hat{v}$ in Eq. (16) is redundant.

The evolution of $f^{1D}_\chi(v)$ can then be obtained by replacing $f_\chi(\vec{v})$ with $f^{1D}_\chi/(4\pi v^2)$ in the Boltzmann equation. Using the fact that $v \propto 1/a$ along free trajectories (where $a$ is the scale factor), and focusing on DM-baryon interactions only, we arrive at

$$a \frac{d}{dt}\bigg|_{\text{free}} [a^{-1} f^{1D}_\chi(v)] = C^{1D}_{\chi s}[f^{1D}_\chi](v), \quad (17)$$

where the 1-dimensional collision operator is

$$C^{1D}_{\chi s}[f^{1D}_\chi](v) = \int dv' \Big[ f^{1D}_\chi(v') \Gamma^{1D}_{\chi s}(v' \to v) \\ - f^{1D}_\chi(v) \Gamma^{1D}_{\chi s}(v \to v') \Big]. \quad (18)$$

### 3. Background heat-exchange rate

Upon solving for the evolution of $f^{1D}_\chi(v)$, one may obtain the background heat-exchange rate from

$$\dot{\mathcal{Q}}_\chi = \frac{1}{2} \rho_\chi \int dv \ v^2 \ C^{1D}_{\chi s}[f^{1D}_\chi](v). \quad (19)$$

For the specific case of a cross-section[4] scaling as $\sigma_{\chi s}(v) = \sigma_n v^n$, this can be rewritten as [42]

$$\dot{\mathcal{Q}}_\chi = c_n \sigma_n \frac{\rho_s \rho_\chi}{M} \left(\frac{T_b}{m_s}\right)^{\frac{n+1}{2}} \int dv \ f^{1D}_\chi(v) \\ \times \left( 3 \frac{T_b}{M} {}_1F_1\left[-\frac{n+3}{2}, \frac{3}{2}, -\frac{m_s}{2T_b}v^2\right] \\ - v^2 \ {}_1F_1\left[-\frac{n+1}{2}, \frac{5}{2}, -\frac{m_s}{2T_b}v^2\right] \right), \quad (20)$$

where ${}_1F_1$ is the confluent hypergeometric function of the first kind, and

$$c_n \equiv \frac{2^{\frac{n+5}{2}}}{3\sqrt{\pi}} \Gamma\left(\frac{n}{2} + 3\right), \quad (21)$$

where $\Gamma$ is the Gamma function. Since the scattering operator vanishes for the equilibrium distribution, one may also obtain $\dot{\mathcal{Q}}_\chi$ by substituting $f^{1D}_\chi(v) \to \Delta f^{1D}_\chi(v) \equiv 4\pi v^2 [\bar{f}_\chi(v) - f^{\text{eq}}_\chi(v)]$ in Eq. (20). This is numerically more robust at early times when DM is close to equilibrium with baryons.

---

[4] More precisely, it is the momentum-exchange cross-section that is relevant in this case.

### D. Fokker-Planck approximation

#### 1. General description

The numerical solution of the exact Boltzmann equation is challenging, as the exact collision operator $C_{\chi s}[f_\chi]$ renders it an integro-differential equation. In Paper-I, a diffusion approximation to the collision operator was derived, in the form of the Fokker-Planck (FP) operator

$$C^{\text{FP}}_{\chi s}[f_\chi](\vec{v}) \equiv \frac{1}{2} \frac{\partial}{\partial v^i} \bigg[ D^{ij}(\vec{v}) \bigg( \frac{\partial}{\partial v^j} f_\chi(\vec{v}) \\ + \frac{m_\chi}{T_b}(v - V_b)^j f_\chi(\vec{v}) \bigg) \bigg], \quad (22)$$

where the symmetric tensor $D^{ij}(\vec{v})$ is a velocity-dependent effective diffusion tensor. In the absence of DM-DM collisions, an approximate solution for the DM distribution, $f^{\text{FP}}_\chi$ is then obtained from solving

$$n_\chi^{-1} \frac{d}{dt}\bigg|_{\text{free}} [n_\chi f^{\text{FP}}_\chi(\vec{v})] = C^{\text{FP}}_{\chi s}[f^{\text{FP}}_\chi](\vec{v}). \quad (23)$$

With this approximate collision operator, the Boltzmann equation becomes a partial differential equation. Upon discretization, its numerical solution involves solving a system of coupled ODEs with a sparse, tri-diagonal coupling matrix, rather than a full coupling matrix as is the case for the exact collision operator.

The FP operator given in Eq. (22) automatically conserves particle number and satisfies the detailed balance relation given by Eq. (14). In addition, the effective diffusion tensor $D^{ij}(\vec{v})$ was constructed in Paper-I so that $C^{\text{FP}}_{\chi s}$ results in exact momentum and heat-exchange rates *for a given DM distribution*. Explicitly, $C^{\text{FP}}_{\chi s}$ would give the same result as the exact collision operator if inserted in Eq. (19) to compute $\dot{\mathcal{Q}}_\chi$, *for any given* $f_\chi(\vec{v})$. However, the DM distribution $f^{\text{FP}}_\chi(\vec{v})$ obtained by solving the Boltzmann-Fokker-Planck equation (23) need not be an accurate estimate of the true DM distribution $f_\chi(\vec{v})$. As a consequence, there is no guarantee that the FP approximation produces accurate heat- and momentum-exchange rates, unless DM-baryon scattering is truly a diffusive process. We now discuss the regime where DM-baryon scattering can be considered diffusive.

#### 2. Expected regime of validity of the FP approximation

DM-baryon scattering can be considered to be diffusive when individual scattering events change the DM velocity by less than the characteristic width $\sqrt{T_\chi/m_\chi}$ of the DM velocity distribution. The change in DM velocity after a scattering event is given by:

$$\Delta \vec{v}_\chi = \frac{m_s}{M} v_{\chi s}(\hat{n}' - \hat{n}), \quad (24)$$

where $v_{\chi s} \equiv |\vec{v}_\chi - \vec{v}_s|$ is the magnitude of the DM-baryon relative velocity (before or after scattering, as it is a conserved quantity), and $\hat{n}, \hat{n}'$ are its directions before and after scattering, respectively. For typical DM and baryon velocities, $v_{\chi s} \sim \sqrt{T_\chi/m_\chi + T_b/m_s}$. Therefore, the magnitude of the velocity change, relative to the characteristic width of the DM distribution, is approximately

$$\frac{|\Delta \vec{v}_\chi|}{\sqrt{T_\chi/m_\chi}} \sim \frac{m_s}{M} \sqrt{1 + \frac{T_b}{T_\chi} \frac{m_\chi}{m_s}}. \quad (25)$$

Since $T_\chi \leq T_b$, we thus find $|\Delta \vec{v}_\chi|/\sqrt{T_\chi/m_\chi} \gtrsim \sqrt{m_s/M}$. Hence, scattering is *not* diffusive for $m_s \gtrsim m_\chi$. When $m_s \ll m_\chi$, scattering is technically diffusive only as long as $T_\chi \gg (m_s/m_\chi) T_b$, which holds until long after DM-baryon thermal decoupling. By the time this condition is no longer met, the DM velocity distribution is much narrower than the baryon thermal velocities, at which point the heat-exchange rate no longer depends on the details of the DM distribution (see Paper-I for a detailed explanation).

From the discussion above, the FP approximation is guaranteed to be accurate for heavy DM particles with mass $m_\chi \gg m_s$ (recalling that this discussion strictly applies to models which start initially coupled and eventually decouple, and not necessarily to models with $n < 0$). Light DM particles ($m_\chi \lesssim m_s$) are difficult to probe with direct-detection experiments, and cosmological probes are thus most useful for this mass regime. However, the accuracy of the FP approximation for $m_\chi \lesssim m_s$ is a priori unknown. The purpose of this work is to quantify its accuracy via comparison against exact results.

### E. Maxwell-Boltzmann approximation

The assumption widely made in the literature is that DM has a MB velocity distribution. In this case, the evolution of $f_\chi$ reduces to solving for its mean velocity and temperature, both of which satisfy *ordinary* differential equations (ODEs). These ODEs can be derived either from the full Boltzmann collision operator, or the FP operator, given that the latter is constructed to give the same momentum and heat-exchange rates, for a given $f_\chi$. Note that these equations are independent of DM-DM scattering, since the DM-DM scattering operator preserves *any* MB distribution.

Specifically for $\sigma_{\chi s}(v) = \sigma_n v^n$, the background evolution of the effective DM temperature under the MB assumption is given by

$$\frac{d}{dt}(a^2 T_\chi^{\rm MB}) = a^2 \, R_n(T_\chi^{\rm MB}) \times (T_b - T_\chi^{\rm MB}), \quad (26)$$

$$R_n(T_\chi) \equiv 2 c_n \sigma_n n_s \frac{m_\chi m_s}{M^2} \left(\frac{T_\chi}{m_\chi} + \frac{T_b}{m_s}\right)^{\frac{n+1}{2}}, (27)$$

where $\sigma_n$ is a constant, and $c_n$ was defined in Eq. (21).

The volumetric heating rate is then simply

$$\dot{\mathcal{Q}}_\chi^{\rm MB} = \frac{3}{2} n_\chi R_n(T_\chi^{\rm MB}) \times (T_b - T_\chi^{\rm MB}). \quad (28)$$

The MB approximation is accurate in the limit that DM self-interactions are efficient at redistributing its velocities towards the maximum-entropy MB distribution, or when DM is tightly coupled to baryons. However, if DM self-interactions are negligible, and/or DM is not in tight thermal contact with baryons, the DM distribution need not be MB. For this reason, we distinguish $T_\chi^{\rm MB}$ – the solution of Eq. (26) – from $T_\chi$, the true DM temperature, defined from its true distribution function through Eq. (3).

### F. Summary of previous results and goal of the present work

In Paper-I, the MB approximation was compared against the FP approximation for the *background* DM velocity distribution and resulting heat-exchange rate, i.e. for a homogeneous and isotropic universe.

First, it was found that the MB approximation and FP solutions are close to one another for $m_s/m_\chi \ll 1$. Given that this is the regime in which the FP approximation is accurate (§II D 2), this implies that the MB approximation is also accurate for $m_s/m_\chi \ll 1$. This agreement between the MB and FP solutions is, however, specific to the background evolution, and somewhat of a mathematical coincidence. Indeed, in the FP approximation, the background DM distribution $\bar{f}_\chi(v)$ satisfies a 1-dimensional diffusion equation. For $m_s/m_\chi \ll 1$, the typical DM velocities are much smaller than the baryon thermal velocities, and as a consequence, the relevant diffusion coefficients are effectively constant over the range of relevant velocities. The solution of a 1-dimensional diffusion equation with constant coefficients is a Gaussian, i.e. a MB distribution. Importantly, and as argued in Paper-I, the MB distribution is an invalid solution in the presence of perturbations (see also [46]).

Second, it was found in Paper-I that the differences between the MB approximation and FP solutions are significant for $m_s \gtrsim m_\chi$, and can grow to order unity for $m_s \gg m_\chi$. In particular, it was found that the MB approximation results in systematic overestimation of the heat-exchange rate, compared to the FP approximation. However, there is no guarantee that the FP solution is itself accurate in this regime.

Our goal is therefore to quantify the accuracy of the FP approximation in the regime $m_s \gtrsim m_\chi$, in which scattering is non-diffusive. The metric of accuracy that we use is the difference between heat-exchange rates calculated in the FP approximation and those obtained with the exact solution of the integro-differential CBE. We now go on to describe our procedure and numerical methods in detail.

## III. EXACT SOLUTION OF THE BACKGROUND COLLISIONAL BOLTZMANN EQUATION

We are interested in quantifying the *maximal* error that the MB approximation induces when computing the heat-exchange rate. As a consequence, we focus on the limiting case where DM self-interactions are completely negligible. We also exclusively consider the background evolution in this paper.

### A. Differential scattering rates

The differential rate at which DM particles of mass $m_\chi$, with initial and final velocities $\vec{v}$ and $\vec{v}'$ respectively, scatter with baryons of mass $m_s$ and velocity distribution $f_s(\vec{v}_s)$ is given by

$$\Gamma_{\chi s}(\vec{v} \to \vec{v}') = n_s \int d^3 v_s f_s(\vec{v}_s) v_{\chi s} \int d^2 \hat{n}' \frac{d\sigma_{\chi s}}{d^2 \hat{n}'} \\ \times \delta^{(3)} \left[ \vec{v}' - \vec{v} - \frac{m_s}{M} v_{\chi s} (\hat{n}' - \hat{n}) \right], \quad (29)$$

where the 3-dimensional Dirac delta function $\delta^{(3)}$ imposes energy and momentum conservation.

Let us now compute the 1-dimensional differential scattering rate $\Gamma_{\chi s}^{1\mathrm{D}}(v \to v')$ defined in Eq. (16). Since $\vec{v}'$ only appears in the Dirac function, we start by factorizing it in terms of its radial and angular parts:

$$\delta^{(3)} \left[ \vec{v}' - \vec{v} - \frac{m_s}{M} v_{\chi s} (\hat{n}' - \hat{n}) \right] \\ = \frac{1}{v'^2} \delta^{(1)} [v' - v_f] \, \delta^{(2)} [\hat{v}' - \hat{v}_f], \quad (30)$$

$$\vec{v}_f \equiv \vec{v} + \frac{m_s}{M} v_{\chi s} (\hat{n}' - \hat{n}). \quad (31)$$

Therefore, we find, upon integrating over $\hat{v}'$ (dropping the redundant $\hat{v}$ integral in Eq. (16)),

$$\Gamma_{\chi s}^{1\mathrm{D}}(v \to v') \equiv n_s \int d^3 v_s f_s(\vec{v}_s) v_{\chi s} \\ \times \int d^2 \hat{n}' \frac{d\sigma_{\chi s}}{d^2 \hat{n}'} \, \delta^{(1)} [v' - v_f]. \quad (32)$$

Note that the final value of this function is independent of $\hat{v}$, even though the integrand depends on it through $v_{\chi s} \equiv |\vec{v} - \vec{v}_s|$ and $v_f$.

For a generic differential scattering cross-section, we see that the calculation of $\Gamma_{\chi s}^{1\mathrm{D}}(v \to v')$ involves an effectively 4-dimensional integral, for each pair of initial and final velocities. However, in the special case where the differential cross-section is isotropic, i.e.

$$\frac{d\sigma_{\chi s}}{d^2 \hat{n}'} = \frac{\sigma_{\chi s}(v_{\chi s})}{4\pi}, \quad (33)$$

we show in Appendix A that $\Gamma_{\chi s}^{1\mathrm{D}}(v \to v')$ can be reduced to a one-dimensional integral.

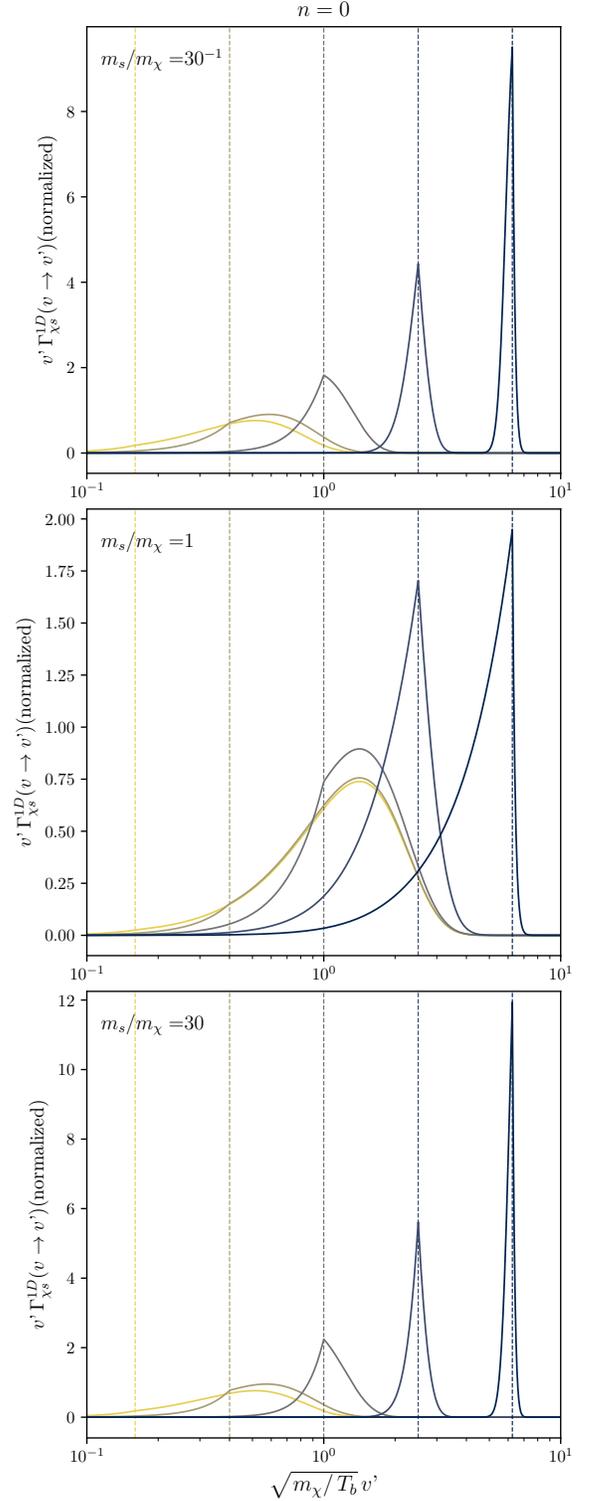

FIG. 1. Differential scattering rates $\Gamma_{\chi s}^{1\mathrm{D}}(v \to v')$ as a function of rescaled $v'$ for $n = 0$ and $m_s/m_\chi = 30^{-1}$ (top panel), $m_s/m_\chi = 1$ (middle), and $m_s/m_\chi = 30$ (bottom). All solid curves are normalized by $\int dv' \Gamma_{\chi s}^{1\mathrm{D}}(v \to v')$. The different colors correspond to different initial velocities, whose rescaled values $\sqrt{m_\chi/T_b}\, v \in \{2.5^{-2}, 2.5^{-1}, 1, 2.5, 2.5^2\}$ are indicated by dashed vertical lines.

In general, the cross-section may have features at characteristic relative velocities (for instance, if the scattering process is mediated by a particle with a finite mass). In this paper, however, we assume a differential cross-section $\sigma_{\chi s}(v_{\chi s}) = \sigma_n v_{\chi s}^n$ with $n \geq 0$, achieved, e.g., through scattering via a sufficiently heavy mediator. Therefore, the only characteristic velocity in Eq. (32) is set by the temperature of baryons. As a consequence, we may factorize the time dependence of $\Gamma_{\chi s}^{1D}(v \to v')$ as follows:

$$\Gamma_{\chi s}^{1D}(v \to v') = \sqrt{\frac{m_\chi}{T_b}} R_n(T_b) \, \widetilde{\Gamma}(u \to u'), \quad (34)$$

$$u \equiv \sqrt{\frac{m_\chi}{T_b}} \, v, \quad (35)$$

where the rate $R_n$ was defined in Eq. (27), and $R_n(T_b) \equiv R_n(T_\chi = T_b)$. The dimensionless coefficients $\widetilde{\Gamma}(u \to u')$ *do not explicitly depend on time*: they only depend on the rescaled DM velocities $u, u'$, as well as on $m_s, m_\chi$ and the index $n$. This factorization allows us to speed up the numerical solution of the Boltzmann equation, as we can pre-compute $\widetilde{\Gamma}(u \to u')$ on a grid of $(u, u')$, and need not recompute it at every timestep. In the case of isotropic differential cross-sections, we have moreover found explicit analytic expressions for $\widetilde{\Gamma}(u \to u')$ for $n = 0, 2, 4$, which we provide in Appendix B. We show the differential scattering rates for $n = 0$ in Fig. 1, for $m_s/m_\chi = 30^{-1}$, 1, 30. Although the reason is made clear later in §IV B, we point out here that for heavy (top panel) and light DM (bottom), the dispersion of the curves around $v' = v$ is smaller than it is for intermediate-mass DM (middle). The differential scattering rates for $n = 2$ and $4$ are qualitatively similar to $n = 0$.

### B. Dimensionless form of the Boltzmann equation

Before proceeding with the numerical methods *per se*, let us first re-write the collisional Boltzmann equation in a way most amenable to efficient numerical integration.

First, given the factorization of the differential scattering rates [Eq. (34)], it is more appropriate to work with the rescaled velocity variable introduced in Eq. (35) and the corresponding rescaled distribution function

$$\widetilde{f}(u) \equiv \sqrt{\frac{T_b}{m_\chi}} f_\chi^{1D}(v). \quad (36)$$

We focus on the redshift range $z \gg 200$, during which the baryon temperature follows closely the radiation temperature, $T_b = T_r \propto 1/a$. As a consequence, the Boltzmann equation for $\widetilde{f}(u)$ becomes

$$a^{1/2} \left. \frac{d}{dt} \right|_{\text{free}} [a^{-1/2} \widetilde{f}(u)] = R_n(T_b) \, \widetilde{C}[\widetilde{f}](u), \quad (37)$$

where the dimensionless collision operator is

$$\widetilde{C}[\widetilde{f}](u) \equiv \int du' \Big[ \widetilde{f}(u') \widetilde{\Gamma}(u' \to u) \\ - \widetilde{f}(u) \widetilde{\Gamma}(u \to u') \Big]. \quad (38)$$

Noting that $u \propto a^{-1/2}$ along free trajectories (assuming, again, that $T_b = T_r \propto 1/a$ at the times of interest), we have

$$\left. \frac{d}{dt} \right|_{\text{free}} = \frac{\partial}{\partial t} - \frac{1}{2} H u \frac{\partial}{\partial u} = H \left( \frac{\partial}{\partial \ln a} - \frac{1}{2} u \frac{\partial}{\partial u} \right), \quad (39)$$

and therefore

$$a^{1/2} \left. \frac{d}{dt} \right|_{\text{free}} [a^{-1/2} \widetilde{f}(u)] = H \left( \frac{\partial \widetilde{f}(u)}{\partial \ln a} - \frac{1}{2} \frac{\partial (u \widetilde{f}(u))}{\partial u} \right), \quad (40)$$

where partial derivatives with respect to $a$ are at constant $u$, and reciprocally.

Equation (37) is a linear and homogeneous integro-differential equation. For positive $n$, it should be solved with initial condition $f_\chi(t_{\text{init}}) = f_\chi^{\text{eq}}$, i.e.

$$\widetilde{f}(u, t_{\text{init}}) = \widetilde{f}^{\text{eq}}(u) \\ \equiv \sqrt{\frac{2}{\pi}} u^2 \exp\left(-\frac{u^2}{2}\right). \quad (41)$$

The collision operator vanishes for the equilibrium distribution $\widetilde{f}^{\text{eq}}(u)$, i.e. $\widetilde{C}[\widetilde{f}^{\text{eq}}](u) = 0$. Using the linearity of $\widetilde{C}$, and the fact that $\partial_a \widetilde{f}^{\text{eq}} = 0$ (at constant $u$), we may therefore rewrite Eq. (37) as the following equation for $\Delta \widetilde{f} \equiv \widetilde{f} - \widetilde{f}^{\text{eq}}$ (using Eq. (40)):

$$\frac{\partial \Delta \widetilde{f}}{\partial \ln a} = \frac{R_n(T_b)}{H} \widetilde{C}[\Delta \widetilde{f}] + \frac{1}{2} \frac{\partial}{\partial u} \left[ u (\widetilde{f}^{\text{eq}} + \Delta \widetilde{f}) \right]. \quad (42)$$

This equation is an *inhomogeneous* integro-differential equation, but with *vanishing initial conditions*, i.e. $\Delta \widetilde{f}(u, t_{\text{init}}) = 0$. A numerical solution of Eq. (42) should therefore be able to extract more accurately the deviations of $f_\chi$ from equilibrium with baryons, which fully determine the heat-exchange rate.

Eq. (42) is valid at $z \gg 200$, during which we may assume that $T_b = T_r \propto 1/a$. This equation can be further simplified if we focus on the radiation-dominated era, during which the expansion rate is given by $H = H_0 \sqrt{\Omega_r} a^{-2}$. Note that this assumption is not required by our approach, but will simply allow us to present results in a more concise way. In this case, we may rewrite Eq. (42) in terms of the variable

$$y \equiv a/a_{\chi b}, \quad (43)$$

where $a_{\chi b}$ is the characteristic thermal decoupling scale factor, defined as [7, 42]

$$a_{\chi b}^{\frac{n+3}{2}} \equiv 2 c_n \sigma_n \frac{n_{s,0}}{H_0 \Omega_r^{1/2}} \left( \frac{M^2}{m_s m_\chi} \right)^{\frac{n-1}{2}} \left( \frac{T_{r,0}}{M} \right)^{\frac{n+1}{2}}, \quad (44)$$

where $n_{s,0} \equiv a^3 n_s$ is the abundance of scatterers at the present time and $T_{r,0} \equiv aT_b$ is the CMB temperature today. In terms of the variable $y$, the collisional Boltzmann equation [Eq. (42)] then takes on the particularly simple form

$$\frac{\partial \Delta \widetilde{f}}{\partial \ln y} = y^{-\frac{n+3}{2}} \widetilde{C}[\Delta \widetilde{f}] + \frac{1}{2}\frac{\partial}{\partial u}\left[u(\tilde{f}^{\text{eq}} + \Delta \widetilde{f})\right]. \quad (45)$$

In the next section we discuss our numerical method for solving this equation.

### C. Numerical implementation of the collisional Boltzmann equation

We solve Eq. (45) over the domain $u \in (u_{\min}, u_{\max}) \equiv (10^{-6}, 8)$. We discretize this range into $N = 1272$ logarithmically-spaced bins, $u_i = u_{\min} e^{i \, d\ln u}$, with $i = 0, ..., N-1$, and $d\ln u = 0.0125$. We also define $\widetilde{f}_i \equiv \widetilde{f}(u_i)$, $\Delta \widetilde{f}_i \equiv \Delta \widetilde{f}(u_i)$, and $\widetilde{\Gamma}_{ij} \equiv \widetilde{\Gamma}(u_i \to u_j)$, for $i,j = 0, ..., N-1$. The discretized form of the collision operator [Eq. (38)] is then

$$\widetilde{C}[\Delta \widetilde{f}](u_i) = \sum_{j=0}^{N-1} M_{ij} \Delta \widetilde{f}_j, \quad (46)$$

$$M_{ij} \equiv d\ln u \left(\widetilde{\Gamma}_{ji} u_j - \delta_{ij} \sum_k \widetilde{\Gamma}_{ik} u_k\right). \quad (47)$$

The matrix $\boldsymbol{M}$ is time invariant and needs to only be computed once at the beginning of the numerical implementation. The gradient operator on the right-hand-side of Eq. (45) is discretized as

$$\left.\frac{\partial(u\widetilde{f})}{\partial u}\right|_i = \frac{1}{2u_i d\ln u}$$
$$\times \begin{cases} (u_0 \widetilde{f}_0 + u_1 \widetilde{f}_1) & i = 0 \\ (u_{i+1}\widetilde{f}_{i+1} - u_{i-1}\widetilde{f}_{i-1}) & 1 \leq i \leq N-2 \\ -(u_{N-2}\widetilde{f}_{N-2} + u_{N-1}\widetilde{f}_{N-1}) & i = N-1 \end{cases}$$
$$\equiv \sum_{j=i-1}^{i+1} \alpha_{ij} \widetilde{f}_j, \quad \text{where} \quad \alpha_{0,-1} = \alpha_{N-1,N} = 0. \quad (48)$$

This implementation is second-order accurate, and enforces number-conservation within machine precision, i.e. $\sum_i u_i d\ln u \, \frac{\partial(u\widetilde{f})}{\partial u}|_i = 0$.

With this discretization, we may transform the integro-partial-differential equation (45) into a system of coupled linear ODEs for the functions $\Delta \widetilde{f}_i(y)$:

$$\frac{\partial \Delta \widetilde{f}_i}{\partial \ln y} = y^{-\frac{n+3}{2}} \sum_{j=0}^{N-1} M_{ij} \Delta \widetilde{f}_j$$
$$+ \frac{1}{2}\sum_{j=i-1}^{i+1} \alpha_{ij}(\widetilde{f}_j^{\text{eq}} + \Delta \widetilde{f}_j). \quad (49)$$

We solve these coupled ODEs using `solve_ivp` in `scipy`'s `integrate` package [47]. Since this system of equations is prone to instabilities at early times, we use the implicit 'BDF' (Backward Differentiation Formula) method which is designed for stiff systems [48–50]. We fix the various parameters for `solve_ivp` as follows. The time range (or `t_span`) is logarithmically spaced, and given by $(\ln y_{\text{init}}, \ln y_{\text{final}}) = (\ln 10^{-2}, \ln 10^3)$. The initial conditions are given by $\Delta \widetilde{f}_i(y_{\text{init}}) = 0$ [Eq. (41)]. The maximum logarithmic $y$-interval (`max_step`) is set to $10^{-2}$, the relative tolerance (`rtol`) is $10^{-3}$, and the absolute tolerance `atol` is set to 0 so that the error $<$ `atol + rtol*abs(`$\Delta \widetilde{f}_i$`)` is dominated by `rtol`.

Finally, after solving for $\Delta \widetilde{f}_i(y)$, we obtain $\dot{\mathcal{Q}}_\chi^{\text{ex}}(y)$ by rewriting Eq. (20) in a dimensionless form as,

$$\frac{\dot{\mathcal{Q}}_\chi^{\text{ex}}}{\frac{3}{2} n_\chi H T_b}(y) = \frac{(m_\chi/M)^{\frac{n-1}{2}}}{3y^{\frac{n+3}{2}}} \int du \, \Delta \widetilde{f}^{\text{ex}}(u,y) \times$$
$$\left[3\frac{m_\chi}{M} \, {}_1F_1\left(-\frac{n+3}{2}, \frac{3}{2}, -\frac{m_s}{2m_\chi}u^2\right)\right.$$
$$\left. -u^2 \, {}_1F_1\left(-\frac{n+1}{2}, \frac{5}{2}, -\frac{m_s}{2m_\chi}u^2\right)\right]. \quad (50)$$

We compute the integral as a simple Riemann sum.

*Convergence and accuracy tests*

We have checked that our results are converged, as $\dot{\mathcal{Q}}_\chi^{\text{ex}}$ changes at most by 0.05% if we halve `max_step`, and at most by 0.02% when we halve $d\ln u$. We performed these convergence tests for $m_s/m_\chi = 30, 1$, and $30^{-1}$ and all $n \in \{0, 2, 4\}$.

Another condition we checked for is that probability (or number of DM particles) is conserved at each time step, for all values of $n$ and $m_s/m_\chi$ that we consider. Explicitly, we found that

$$\frac{\sum_j u_j \, \Delta \widetilde{f}_j(y)}{\sum_j u_j \left|\Delta \widetilde{f}_j(y)\right|} \leq 10^{-5} \quad \forall \, y, \quad (51)$$

for $n = 4$, and in fact is $\lesssim 10^{-7}$ for $n = 2$, and $\lesssim 10^{-8}$ for $n = 0$.

We also checked that our results are converged with respect to `rtol` (for $m_s/m_\chi = 30, 1, 30^{-1}$ and $n \in \{0, 2, 4\}$), and find that $\dot{\mathcal{Q}}_\chi^{\text{ex}}$ changes at most by 0.05% when `rtol` is reduced by a factor of 10.

As an additional sanity check, we have confirmed that our implementation does recover the evolution of $T_\chi^{\text{MB}}$ when we force the DM velocity distribution to be MB. This check was performed as follows. We initialize the computation with the same initial conditions, $\Delta \widetilde{f}_i(y_{\text{init}}) = 0$. After each logarithmic interval of $d\ln y = 10^{-2}$, we compute the DM temperature from

$$T_\chi(y) = T_b(y)\left(1 + \frac{1}{3}\sum_i \Delta \widetilde{f}_i(y) u_i^3 d\ln u\right). \quad (52)$$

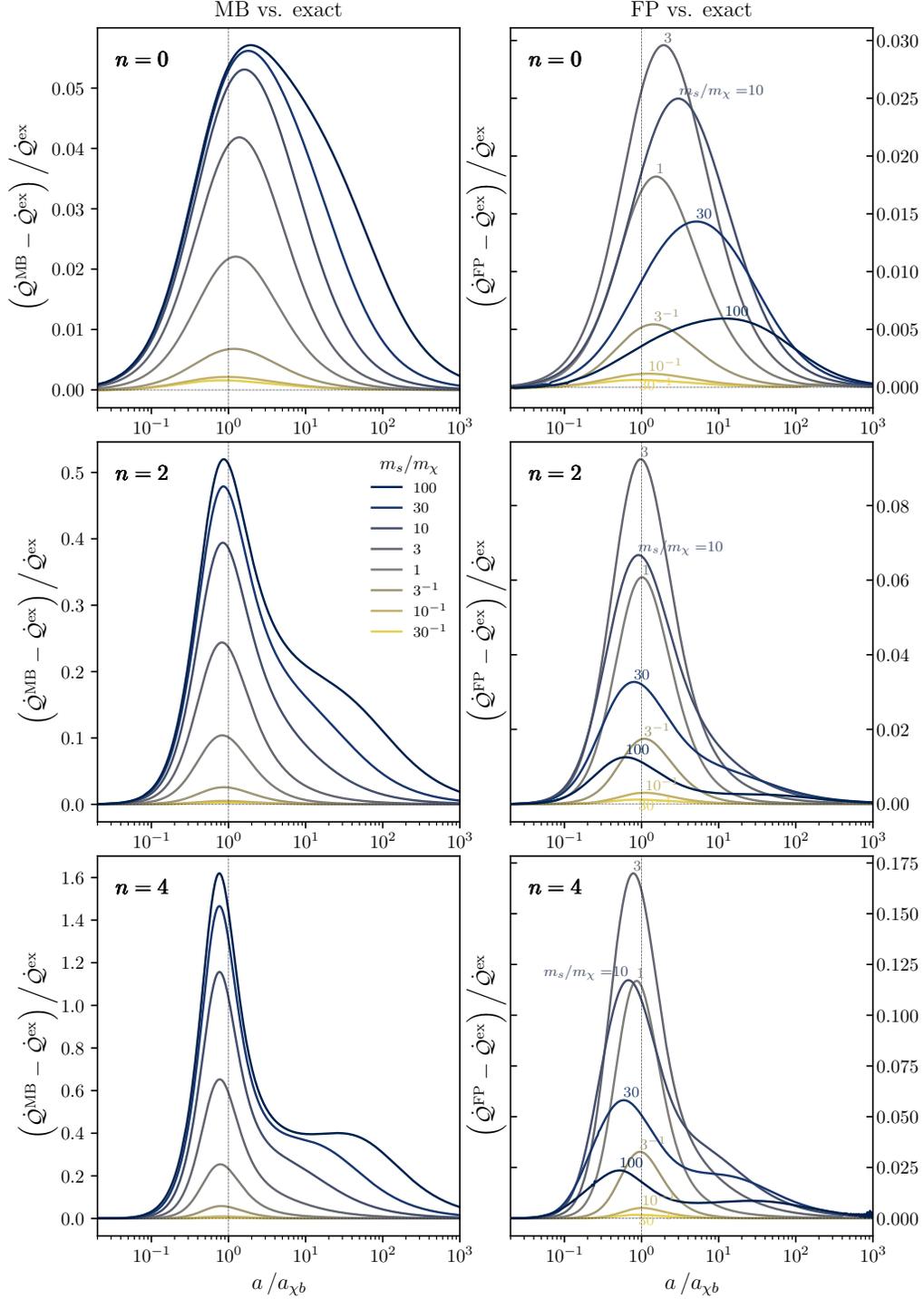

FIG. 2. Relative differences between heat-exchange rates $\dot{\mathcal{Q}}_\chi$ obtained via different methods discussed in this paper, as a function of the scale factor, $a$, normalized by the DM-baryon thermal decoupling time, $a_{\chi b}$ [Eq. (44)]. Each row is for a particular value of $n$, and each colored curve represents a particular mass ratio $30^{-1} \leq m_s/m_\chi \leq 10^2$. The vertical line marks $a = a_{\chi b}$. **Left:** The fractional difference between $\dot{\mathcal{Q}}_\chi$ obtained using the MB assumption [Eq. (28)] and our exact implementation [Eq. (20)]. These results are similar to the relative difference between $\dot{\mathcal{Q}}_\chi^{\rm MB}$ and $\dot{\mathcal{Q}}_\chi^{\rm FP}$ shown in Figure 3 of Paper-I. **Right:** The fractional difference between $\dot{\mathcal{Q}}_\chi$ obtained using the FP approximation (taken from Paper-I) and our exact method. The FP method gives highly accurate background heat-exchange rates when $m_s/m_\chi \ll 1$ and $m_s/m_\chi \gg 1$. We see that the largest error in $\dot{\mathcal{Q}}_\chi^{\rm FP}$ (reached for $m_s/m_\chi \sim 3$) is lower than the largest error in $\dot{\mathcal{Q}}_\chi^{\rm MB}$ by a factor of $\sim 2$ for $n=0$ (top row), and by as much as a factor of $\sim 10$ for $n=4$ (bottom row).

Then we enforce $\widetilde{f}(y)$ to be MB-distributed at the temperature $T_\chi(y)$, i.e. set

$$\Delta \widetilde{f}_i(y) = \widetilde{f}_i^{\text{MB}}(y)\big|_{T_\chi} - \widetilde{f}_i^{\text{eq}}, \quad (53)$$

$$\widetilde{f}_i^{\text{MB}}(y)\big|_{T_\chi} \equiv \sqrt{\frac{2}{\pi}}\left(\frac{T_b}{T_\chi}\right)^{\frac{3}{2}} u_i^2 \exp\left[-\frac{T_b}{T_\chi}\frac{1}{2}u_i^2\right]. \quad (54)$$

This step is equivalent to enforcing strong DM self-interactions, which reshuffle DM velocities into the maximum-entropy distribution. Lastly, we input this new $\Delta \widetilde{f}_i(y)$ into solve_ivp as the initial condition for the next logarithmic $y$-step, and iterate until $y_{\text{final}}$. We perform this procedure for $n = 2$, $m_s/m_\chi \in \{30^{-1}, 1, 30\}$, extract the evolution of $T_\chi(y)$, and compare it to $T_\chi^{\text{MB}}$ (the latter is obtained by numerically solving Eq. (26) using solve_ivp with the same parameters as prescribed in the last section). We find that the fractional difference between the two is always $\leq 0.7\%$.

## IV. RESULTS

### A. Description of the results

We solve the collisional Boltzmann equation for $\overline{f}_\chi(v)$ as described in §III C, and compute the corresponding heat-exchange rate $\dot{\mathcal{Q}}_\chi^{\text{ex}}$, for a range of mass ratios $30^{-1} \leq m_s/m_\chi \leq 10^2$, and for $n \in \{0, 2, 4\}$.

We also compute the corresponding quantities $\overline{f}_\chi^{\text{FP}}(v)$ and $\dot{\mathcal{Q}}_\chi^{\text{FP}}$ within the Fokker-Planck approximation, using the code described in Paper-I. Lastly, we compute $\dot{\mathcal{Q}}_\chi^{\text{MB}}$ by rewriting Eq. (28) in a dimensionless form with variables $X^{\text{MB}} \equiv T_\chi^{\text{MB}}/T_b$ and $y$ as (following Paper-I),

$$\frac{\dot{\mathcal{Q}}_\chi^{\text{MB}}}{\frac{3}{2}n_\chi H T_b} = \frac{d}{dy}\left(y X^{\text{MB}}\right)$$
$$= \left(\frac{1 + (m_s/m_\chi)X^{\text{MB}}}{1 + m_s/m_\chi}\right)^{\frac{n+1}{2}} \frac{1 - X^{\text{MB}}}{y^{\frac{n+3}{2}}} \quad (55)$$

with the initial condition $X^{\text{MB}}(y_{\text{init}}) = 1$. As mentioned earlier, DM-baryon scattering in models with $n \geq 0$ decouples well before recombination, and so our analysis is valid for $z \gg z_{\text{rec}}$.

We show the fractional differences between $\dot{\mathcal{Q}}_\chi^{\text{MB}}$ and the exact $\dot{\mathcal{Q}}_\chi^{\text{ex}}$ in the left column of Fig. 2. These look qualitatively and quantitatively similar to Fig. 3 in Paper-I, where the relative difference between $\dot{\mathcal{Q}}_\chi^{\text{MB}}$ and $\dot{\mathcal{Q}}_\chi^{\text{FP}}$ is shown: the fractional difference is small for $m_s/m_\chi \lesssim 1$, and increases with $m_s/m_\chi$, reaching order unity for $m_s \gg m_\chi$ and for indices $n \geq 2$.

The right column of Fig. 2 explains this similarity: there, we show the relative difference between $\dot{\mathcal{Q}}_\chi^{\text{FP}}$ and $\dot{\mathcal{Q}}_\chi^{\text{ex}}$. We see that this fractional difference never exceeds 17% across all mass ratios and power-law indices considered. This indicates that the FP approximation fares well even in the regime where DM-baryon scattering is not, a priori, diffusive.

Interestingly, the right panel of Fig. 2 shows that the FP approximation is most accurate not only in the regime $m_s \ll m_\chi$, where this accuracy is expected, but also in the opposite regime $m_s \gg m_\chi$, where, naively, we would have expected it to be the least accurate. Instead, our results indicate that the FP approximation is least accurate for intermediate mass ratios, $m_s/m_\chi \sim$ few.

This surprising result is not the outcome of chance cancellations in the heat-exchange rate, but rather comes from the accuracy of $\overline{f}_\chi^{\text{FP}}(v)$ itself. We demonstrate this in Fig. 3, where we explicitly show $\overline{f}_\chi^{\text{ex}}$ and its deviation from $\overline{f}_\chi^{\text{FP}}$ for $m_s \ll m_\chi$ (left column), $m_\chi = m_s$ (middle column), and $m_s \gg m_\chi$ (right column) for each value of $n$. We see that the FP approximation produces an accurate DM distribution function ($|\overline{f}_\chi^{\text{FP}} - \overline{f}_\chi^{\text{ex}}| \ll 1$), not only for $m_s \ll m_\chi$ as expected, but for $m_s \gg m_\chi$ as well. When $m_s \ll m_\chi$, the exact distribution (blue-purple curves) stays close to the MB solution (dotted black curve). In the regime where $m_s \gg m_\chi$, the exact distribution deviates significantly from MB, but is well recovered by the FP approximation. The largest deviations $|\overline{f}_\chi^{\text{FP}} - \overline{f}_\chi^{\text{ex}}| \sim \mathcal{O}(10-20\%)$ occur for the intermediate mass ratio $m_s \sim m_\chi$.

Before discussing the mass dependence of the results in Sec. IV B below, let us quickly comment on the time-dependence of the relative differences shown in Fig. 2. We can see that both the MB and FP approximations are typically least accurate around the time of thermal decoupling $a \sim a_{\chi b}$ (broadly speaking), although with different maximum errors, and with a different mass dependence. This can be understood quite simply: well before decoupling, the heat-exchange rate approaches the quasi-steady-state value $\dot{\mathcal{Q}}_\chi \to \frac{3}{2}n_\chi H T_b$, regardless of the method; well after decoupling, the heat-exchange rate becomes independent of the details of DM velocity distribution, and thus of the method used to compute it, as discussed in Paper-I.

### B. Understanding the results

First, let us understand the reason behind the accuracy of the FP solution in the regime $m_s \gg m_\chi$. In §II D 2, we determined the regime of validity of the diffusion approximation by considering the typical *magnitude of the velocity vector change* per scattering, $|\Delta \vec{v}| \equiv |\vec{v}' - \vec{v}|$. This is an accurate indicator for *3-dimensional* diffusion in velocity (vector) space. However, when focusing on the background DM distribution, what is relevant is, instead, the typical *change in velocity magnitude* per scattering, $|\Delta v| \equiv |v' - v|$. If this change is small relative to the characteristic width of $\overline{f}_\chi(v)$, then DM-baryon scattering is indeed diffusive in the *1-dimensional* space of velocity magnitudes. Given that $|\Delta v| \leq |\Delta \vec{v}|$ by the triangle inequality, it is possible for DM-b scattering to be diffusive in one dimension even if it is not diffusive in three dimensions. Let us estimate $|\Delta v|$ for $m_s \gg m_\chi$. Again, we start from the change in DM velocity during a scattering event:

$$\vec{v}' = \vec{v} + \frac{m_s}{M} v_{\chi s}(\hat{n}' - \hat{n}) = \vec{v} + \frac{m_s}{M}(v_{\chi s}\hat{n}' - \vec{v}_{\chi s}), \quad (56)$$

where we recall that $\vec{v}_{\chi s} \equiv \vec{v} - \vec{v}_s$ is the initial DM-baryon relative velocity. The ratio of the charactersitic baryon and DM velocities is $v_s/v_\chi \sim \sqrt{(T_b/T_\chi)/(m_s/m_\chi)}$. Therefore, if $m_s \gg m_\chi$, and as long as $T_b/T_\chi \ll m_s/m_\chi$, we find that $v_s \ll v_\chi$. Thus, for typical velocities,

$$\vec{v}_{\chi s} = \vec{v}\left(1 + \mathcal{O}\left(\sqrt{(T_b/T_\chi)/(m_s/m_\chi)}\right)\right). \quad (57)$$

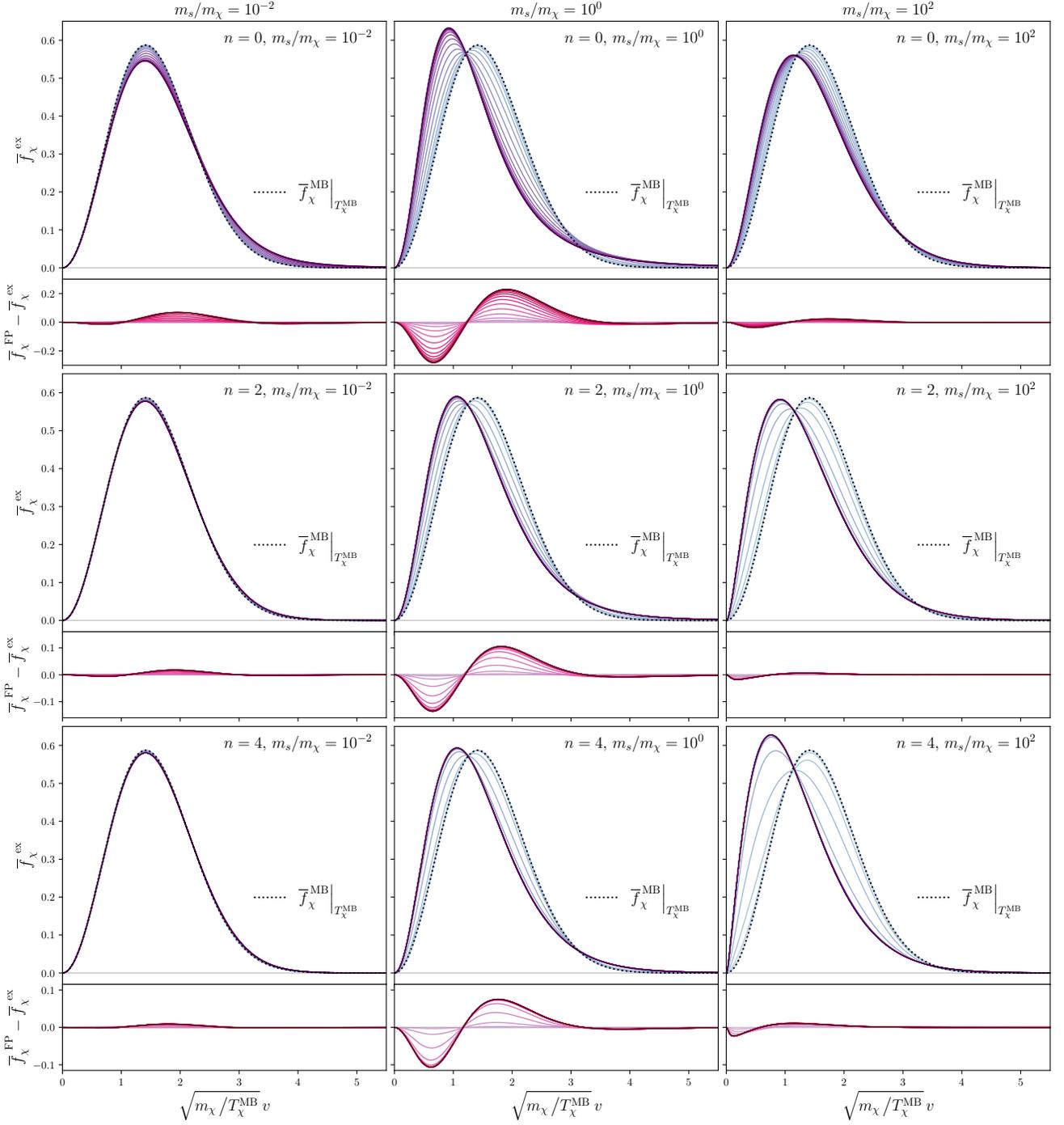

FIG. 3. Visualization of our numerical results, showing the time-evolution of $\overline{f}_\chi^{\rm ex}$, plotted as a function of $\sqrt{m_\chi/T_\chi^{\rm MB}}\, v$ for different $m_s/m_\chi \in \{10^{-2}, 1, 10^2\}$ (increasing from left to right) and different $n \in \{0, 2, 4\}$ (increasing from top to bottom). Each panel comprises two subplots: the top subplot of $\overline{f}_\chi^{\rm ex}$ (blue-purple curves) and a smaller bottom subplot showing deviations of the FP solution from the exact one, $\overline{f}_\chi^{\rm FP} - \overline{f}_\chi^{\rm ex}$ (pink curves). The progression from light-colored curves to darker ones represents time-evolution from initial $a_{\rm init}/a_{\chi b} = 10^{-2}$ to final $a_{\rm final}/a_{\chi b} = 10^3$. The black dotted curve (amidst the $\overline{f}_\chi^{\rm ex}$ curves) is the MB distribution $\overline{f}_\chi^{\rm MB}$ at temperature $T_\chi^{\rm MB}$. As explained in §IV A–IV B, both the FP and MB solutions approximate the exact background distribution well for $m_s \ll m_\chi$ (leftmost column), but the FP approximation closely tracks the exact distribution even for $m_s \gg m_\chi$ (rightmost column) – when the MB solution is highly inaccurate.

Substituting $m_s/M = 1 - m_\chi/M = 1 + \mathcal{O}(m_\chi/m_s)$ in Eq. (56), we see that the term $-\vec{v}_{\chi s}$ nearly cancels the term $\vec{v}$, and we are left with

$$\vec{v}' \approx v\,\hat{n}', \quad (58)$$

up to relative corrections of order $\sqrt{T_b m_\chi/T_\chi m_s}$. Therefore, we see that in the regime $m_s \gg m_\chi$, and as long as $T_b/T_\chi \ll m_s/m_\chi$, elastic scattering merely changes the directions of DM velocities, without changing their magnitudes. That is, $|v' - v| \ll v_\chi$, even though $|\vec{v}' - \vec{v}| \sim v_\chi$ is not small[5]. This result can be seen in Fig. 1, which shows that the differential scattering rates $\Gamma^{1D}_{\chi s}(v \to v')$ are narrowly distributed around $v' \approx v$ for $m_s \ll m_\chi$ as well as for $m_s \gg m_\chi$, while they are broad for $m_s \sim m_\chi$.

To make this point more quantitative, in Appendix C we compute the following dimensionless quantities:

$$\Delta^2_{3D}(v) \equiv \frac{\langle \sigma_{\chi s} v_{\chi s}(\Delta \vec{v})^2\rangle}{\langle \sigma_{\chi s} v_{\chi s}\rangle v^2}, \quad (59)$$

$$\Delta^2_{1D}(v) \equiv \frac{\langle \sigma_{\chi s} v_{\chi s}\,\Delta(v^2)\rangle}{\langle \sigma_{\chi s} v_{\chi s}\rangle v^2}. \quad (60)$$

These quantities determine whether DM-baryon scattering is a diffusive process in three dimensions and one dimension, respectively. We show $\Delta^2_{1D}$ and $\Delta^2_{3D}$ evaluated at $v_\chi = \sqrt{T_b/m_\chi}$ in Fig. 4, as a function of $m_s/m_\chi$, and for $n \in \{0, 2, 4\}$. We see that $\Delta^2_{1D} \leq \Delta^2_{3D}$ for all mass ratios, as expected from the triangle inequality. We see that both quantities are small for $m_s/m_\chi \ll 1$, indicating that scattering is diffusive in both one and three dimensions for heavy DM. For $m_s/m_\chi \sim 1$, both quantities are larger than unity, and increasingly so with larger indices $n$, indicating that scattering is non-diffusive in both one and three dimensions for comparable DM-baryon masses.

---

[5] In this regime, one can think of DM particles as fast ping-pong balls, and baryons as slow-moving billiard balls: upon scattering with the billiard balls, the ping-pong balls change their direction of motion with very little change to their velocity magnitude.

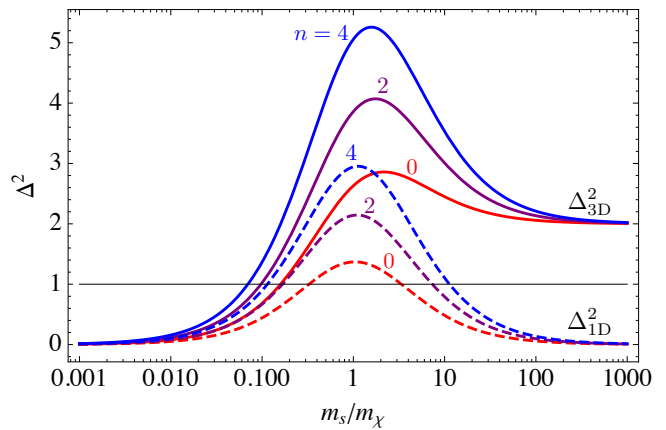

FIG. 4. Dimensionless amplitude of 3-dimensional (solid lines) and 1-dimensional (dashed lines) DM velocity diffusion, as a function of scatterer-to-DM mass ratio $m_s/m_\chi$, evaluated at $v_\chi = \sqrt{T_b/m_\chi}$, and for power-law indices $n = 0$ (red), $n = 2$ (purple) and $n = 4$ (blue). Having $\Delta^2_{3D} \ll 1$ indicates diffusive DM-baryon scattering in 3-dimensional velocity (vector) space, and $\Delta^2_{1D} \ll 1$ indicates diffusive DM-baryon scattering in 1-dimensional velocity (magnitude) space.

Lastly, for $m_s/m_\chi \gg 1$, $\Delta^2_{1D} \ll 1$ and $\Delta^2_{3D} \to 2$, indicating that scattering is diffusive in one dimension but not in three dimensions.

These considerations explain, a posteriori, why the FP approximation is most accurate for $m_s \ll m_\chi$ as well as $m_s \gg m_\chi$ for the *background* DM distribution and heat-exchange rate. Interestingly, even in the regime where $m_s \sim m_\chi$, and DM-baryon scattering is non-diffusive in both one and three dimensions (especially so for $n = 4$), the FP approximation gets the heat-exchange rate within $\sim 20\%$ accuracy for $n = 4$, within $\sim 10\%$ for $n = 2$, and within a few percent for $n = 0$. This bodes well for its accuracy when treating inhomogeneities and anisotropies.

Our focus in this paper is on the accuracy of the FP approximation, but it is worth recalling in which regimes the MB approximation is accurate, and why that is. As we discussed above, there is a well-defined *physical* regime of accuracy for the FP approximation, which is when scattering is diffusive (be it in 1D or 3D). In contrast, regimes in which the MB approximation is accurate do not arise from the satisfaction of a physical criterion, but rather from a interesting and purely *mathematical coincidence*, which is that the solution of a 1D diffusion equation with a uniform (velocity-independent) diffusion coefficient is a Gaussian (see Paper-I and [46]). This explains why the *background* MB approximation is closest to the FP solution (thus to the exact result) for $m_s \ll m_\chi$, since the diffusion coefficient is nearly independent of $v_\chi$ across the width of the DM velocity distribution in this regime. It also explains why the *background* MB approximation is rather accurate for all masses for $n = 0$, and decreasingly accurate with increasing $n$; indeed, the diffusion coefficients become steeper functions of DM velocity with increasing $n$ – had we considered the case $n = -1$, we would have found that the background MB solution is exactly equal to the FP solution for all masses, as the diffusion coefficients are strictly constant in that case. As discussed in Paper-I, such a mathematical coincidence does not carry over to the anisotropic diffusion equation. Therefore, there is no guarantee that the perturbed MB approximation is accurate for *any* mass ratio and for *any* index $n$. In particular, our results for the *background* heat-exchange rate should not be taken as a hint that the MB momentum-exchange rate might be accurate beyond the order-unity level in any situation: it probably is not, even for $n = 0$, and even for $m_s \ll m_\chi$.

## V. CONCLUSIONS

We have quantified the accuracy of the Fokker-Planck (FP) approximation of the Boltzmann collision operator for elastic DM-baryon scattering in a homogeneous and isotropic background. The models we studied in this work have an isotropic differential scattering cross-section proportional to an even and positive power in DM-baryon relative velocity ($d\sigma_{\chi s}/d^2\hat{n}' \propto v_{\chi s}^n$, $n \geq 0$). We worked in the limit where DM-DM scattering is negligible, and thus found the *maximal* error in the DM-baryon heat-exchange rate induced by the FP approximation. We have additionally determined the maximal error arising from assuming a thermalized or Maxwell-Boltzmann (MB) DM velocity distribution. The MB assumption is ubiquitously used in the literature and is the limiting regime where either DM-baryon or DM-DM scattering ensues with perfect efficiency at any given time.

Considering an isotropic differential DM-baryon scattering cross-section for simplicity, we have reduced the differential scattering rate—which is generally a 4D integral—to a 1D integral. Furthermore, we have found fully analytical expressions for the differential scattering rate in models where the differential cross-section is a positive and even-integer power law in the DM-baryon relative velocity.

In order to quantify accuracy, we compared background heat-exchange rates obtained from the FP and MB methods with those obtained from our exact implementation. The errors arising in the FP method are no more than 3%, 10%, and 17%, for $n = 0, 2, 4$, respectively, in contrast with errors of up to 6%, 50%, and 160%, for the MB approximation. This significant improvement is a positive indicator for the accuracy of the FP approximation in the presence of anisotropies and inhomogeneities as well, where one can then extract a potentially more accurate momentum-exchange rate than that within the MB assumption. Reassuringly, maximum errors of order unity in the MB heat-exchange rate suggest similar errors for the momentum-exchange rate and cosmological upper limits on $\sigma_n$. Nevertheless, the non-negligible differences between the exact and MB heat-exchange rates (especially for $n \geq 2$) indicate that current analyses are missing non-negligible aspects of DM-baryon scattering. Besides, as we argued here and in Paper-I, the MB approximation is sometimes accurate for the *background* solution (especially for $n = 0$) due to a mathematical property of the solution of *one-dimensional* diffusion equations; however this property does not carry over to three dimensions, and the MB approximation is likely inaccurate at the order-unity level for the momentum-exchange rate.

Our quantitative results strictly only apply to models with an isotropic DM-baryon differential cross-section. However, the qualitative conclusions should remain unchanged for arbitrary cross-sections, which will require more involved numerical evaluation of the differential scattering rate. For simplicity, we have also assumed a radiation-dominated universe throughout our calculations (which is valid for the models we study, given current upper limits on the cross-sections), but the modification to a generic expansion rate is straightforward and we have checked that it does not affect our results in any significant way. We limit our study to the period of photon-baryon thermal coupling, valid up until redshift $z \sim 200$. Since the models we consider undergo DM-baryon decoupling when $z \gg 200$ (deep in the radiation-dominated era), this assumption does not affect our results.

The most significant limitation of this work is that it only applies to background quantities and cannot straightforwardly be extrapolated to perturbations, relevant to CMB-anisotropy and large-scale structure observations. In particular, one cannot use our reported fractional errors in MB and FP heat-exchange rates to estimate the errors in the MB and FP momentum-exchange rates with respect to the corresponding rate obtained from an exact calculation[6]. It is reasonable to expect that the accuracy of momentum-exchange rates should mirror that of the heat-exchange rates. In particular, we can expect that the Boltzmann-FP hierarchy derived in Ref. [42] should provide significantly more accurate momentum-exchange rates than those traditionally obtained within the MB assumption. This expectation is reinforced by the fact that, for the background evolution, the FP approximation remains quite accurate even in the regime where scattering is physically non-diffusive. Still, for a rigorous confirmation of this projection, one would have to implement and solve the exact collisional Boltzmann equation, accounting for inhomogeneities and perturbations, a task well beyond the scope of this study. A more tractable next step may be to compare the exact and FP solutions of a simplified anisotropic problem, for instance one in which the baryon-DM bulk relative velocity is imposed rather than being solved self-consistently.

---

[6] It is well known that, if the DM has a MB distribution, the heat- and momentum-exchange rates are proportional to the difference of DM-baryon temperatures and bulk velocities, respectively, and that the proportionality coefficients are related by simple mass ratios. Such simple relationships no longer hold when one lifts the MB assumption.

Ultimately, a comprehensive analysis will include *both* DM-baryon and DM-DM scattering as prescribed by the respective model-dependent cross-sections. This would require implementing a non-linear collision operator in conjunction with the evolution equations for perturbations. It is in this highly non-trivial scenario that the Fokker-Planck approximation can potentially bring about significant simplifications, while producing more accurate results than current methods. Once this full framework is implemented, it will open up new windows into the microphysical properties of dark matter. Next-generation missions such as the Rubin Observatory [17, 51], the Simons Observatory [45] and CMB-Stage IV experiments [39] are forecasted to be sensitive to DM-baryon cross sections about 2, 8, and 26 times smaller, respectively, than current *Planck* limits, promising at the very least much tighter limits, as well as the tantalizing possibility of a detection. This makes it all the more critical that the standard formalism be generalized to be able to account for DM self-scattering. The methods and results presented in this work are important steps towards achieving this goal.

### ACKNOWLEDGEMENTS

We thank Kimberly K. Boddy, Nils Schöneberg, Nanoom Lee, Trey W. Jensen, and Vera Gluscevic for detailed comments that helped improve this manuscript. This work is supported by NASA grant No. 80NSSC20K0532. YAH is a CIFAR-Azrieli Global scholar and also acknowledges support from NSF grant No. 1820861. SSG is funded by the James Arthur Graduate Associate fellowship.

### Appendix A: Simplified expressions for $\Gamma_{\chi s}^{\rm 1D}(v \to v')$ in the case of isotropic scattering

The goal of this appendix is to derive analytic expressions for the angle-integrated differential scattering rate $\Gamma_{\chi s}^{\rm 1D}(v \to v')$ given in Eq. (32), in the case of isotropic differential scattering cross-section, i.e. for $d\sigma_{\chi s}/d^2\hat{n}' = \sigma_{\chi s}(v_{\chi s})/4\pi$.

We start by rewriting the Dirac delta as

$$\delta^{(1)}[v' - v_f] = 2v'\delta^{(1)}[v'^2 - v_f^2]. \quad (A1)$$

We then rewrite the vector $\vec{v}_f$ defined in Eq. (31) as

$$\vec{v}_f = \frac{m_s}{M}v_{\chi s}\hat{n}' + \vec{x}, \quad (A2)$$

$$\vec{x} \equiv \vec{v} - \frac{m_s}{M}\vec{v}_{\chi s}, \quad (A3)$$

and rewrite the argument of the Dirac delta as follows:

$$v'^2 - v_f^2 = 2\frac{m_s}{M}v_{\chi s}x \ (X - \hat{n}' \cdot \hat{x}),$$

$$X \equiv \frac{v'^2 - x^2 - \left(\frac{m_s}{M}v_{\chi s}\right)^2}{2\frac{m_s}{M}v_{\chi s}x}.$$

We therefore have isolated the dependence in $\hat{n}'$ as follows

$$\delta^{(1)}[v' - v_f] = \frac{M}{m_s}\frac{v'}{v_{\chi s}}\frac{1}{x}\delta^{(1)}[X - \hat{n}' \cdot \hat{x}]. \quad (A4)$$

Under the assumption that the differential cross-section is isotropic, we may therefore compute the innermost integral in Eq. (32):

$$\mathcal{I} \equiv \int d^2\hat{n}' \frac{d\sigma_{\chi s}}{d^2\hat{n}'} \ \delta^{(1)}[v' - v_f] = \sigma_{\chi s}(v_{\chi s})\frac{M}{2m_s}\frac{v'}{v_{\chi s}}$$
$$\times \frac{1}{x}\Theta(1 - X^2), \text{(A5)}$$

where $\Theta$ is the Heaviside step function imposing that the integral is non-zero only when $|X| < 1$. A straightforward analysis shows that this condition is satisfied provided

$$|v' - (m_s/M)v_{\chi s}| < x < v' + (m_s/M)v_{\chi s}. \quad (A6)$$

Moreover, using the definition of $\vec{x}$ [Eq. (A3)] we have

$$x = \left| v^2 + \left(\frac{m_s}{M}v_{\chi s}\right)^2 - 2\frac{m_s}{M}v_{\chi s}v\,\mu \right|^{1/2}$$
$$\Rightarrow \mu = \frac{v^2 + \left(\frac{m_s}{M}v_{\chi s}\right)^2 - x^2}{2\frac{m_s}{M}v_{\chi s}v}, \quad (A7)$$

where $\mu \equiv \hat{v} \cdot \hat{n}$. Imposing $|\mu| < 1$ gives additional limits on $x$,

$$|v - (m_s/M)v_{\chi s}| < x < v + (m_s/M)v_{\chi s}. \quad (A8)$$

Hence, we find that

$$\mathcal{I} = \sigma_{\chi s}(v_{\chi s})\frac{M}{2m_s}\frac{v'}{v_{\chi s}}\frac{1}{x} \quad \text{if} \quad x_1 < x < x_2, \text{(A9)}$$

and vanishes otherwise, where

$$x_1 \equiv \max\left(\left|v - \frac{m_s}{M}v_{\chi s}\right|, \left|v' - \frac{m_s}{M}v_{\chi s}\right|\right),$$
$$x_2 \equiv \min(v, v') + \frac{m_s}{M}v_{\chi s}. \quad (A10)$$

The condition $x_1 < x_2$ imposes a lower bound on $v_{\chi s}$:

$$v_{\chi s} > \frac{M}{2m_s}|v' - v|. \quad (A11)$$

We may therefore rewrite Eq. (32) as

$$\Gamma_{\chi s}^{\rm 1D}(v \to v') = \frac{n_s M}{2m_s}v' \int d^3\vec{v}_{\chi s}f_s(|\vec{v}_{\chi s} - \vec{v}|)\sigma_{\chi s}(v_{\chi s})$$
$$\times \frac{1}{x}\,\Theta\left(v_{\chi s} - \frac{M}{2m_s}|v' - v|\right)\Theta(x - x_1)\Theta(x_2 - x), \text{(A12)}$$

where we changed integration variables from $\vec{v}_s$ to $\vec{v}_{\chi s}$. Next, we evaluate the $\vec{v}_{\chi s}$ integral by orienting the polar vector along $\hat{v}$, with $\mu = \hat{v} \cdot \hat{n} = \hat{v} \cdot \hat{v}_{\chi s}$. Then $\int d^3 v_{\chi s} \to 2\pi \int dv_{\chi s} v_{\chi s}^2 \int_{-1}^{1} d\mu$. We change integration variables $\mu \to x$, with [Eq. (A7)]

$$d\mu = \frac{M}{m_s v_{\chi s} v} x\, dx. \quad (A13)$$

Lastly, we recall that baryons have a MB velocity distribution, which we rewrite in terms of $v_{\chi s}$ and $x$ variables as follows:

$$\begin{aligned}
&(2\pi T_b/m_s)^{3/2} f_s^{\mathrm{MB}}(|\vec{v}_{\chi s} - \vec{v}|) \\
&= \exp\left(-\frac{m_s}{2T_b}|\vec{v} - \vec{v}_{\chi s}|^2\right) \\
&= \exp\left(\frac{m_\chi}{2T_b} v^2 - \frac{M}{2T_b} x^2 - \frac{m_s m_\chi}{2MT_b} v_{\chi s}^2\right). \quad (A14)
\end{aligned}$$

Combining everything, we obtain

$$\begin{aligned}
&\Gamma_{\chi s}^{\mathrm{1D}}(v \to v') \\
&= \frac{n_s}{\sqrt{8\pi}\,(T_b/m_s)^{3/2}} \left(\frac{M}{m_s}\right)^2 \frac{v'}{v} \exp\left[\frac{m_\chi}{2T_b} v^2\right] \\
&\quad \times \int_{\frac{|v'-v|}{2m_s/M}}^{\infty} dv_{\chi s}\, v_{\chi s} \sigma_{\chi s}(v_{\chi s}) \exp\left[-\frac{m_s m_\chi}{2MT_b} v_{\chi s}^2\right] \\
&\quad \times \int_{x_1}^{x_2} dx \exp\left[-\frac{M}{2T_b} x^2\right]. \quad (A15)
\end{aligned}$$

The $x$-integral is analytic, and we thus arrive at

$$\begin{aligned}
\Gamma_{\chi s}^{\mathrm{1D}}(v \to v') &= \frac{n_s}{4} \sqrt{\frac{M}{m_s}} \frac{M}{T_b} \frac{v'}{v}\, \exp\left[\frac{m_\chi}{2T_b} v^2\right] \\
&\quad \times \gamma(v, v'), \quad (A16)
\end{aligned}$$

where the function $\gamma(v, v')$ is symmetric in initial and final velocities, and defined as

$$\begin{aligned}
\gamma(v, v') &\equiv \int_{\frac{M|v'-v|}{2m_s}}^{\infty} dv_{\chi s} v_{\chi s} \sigma_{\chi s}(v_{\chi s}) \exp\left[-\frac{m_s m_\chi}{2MT_b} v_{\chi s}^2\right] \\
&\quad \times \left(\mathrm{erf}\left[\sqrt{\frac{M}{2T_b}} x_2\right] - \mathrm{erf}\left[\sqrt{\frac{M}{2T_b}} x_1\right]\right). \quad (A17)
\end{aligned}$$

Given that $\gamma(v, v')$ is symmetric in $v \leftrightarrow v'$, we only need to calculate it for $v \leq v'$. In that case, $x_2 = v + (m_s/M) v_{\chi s}$, independent of $v_{\chi s}$. To determine $x_1$, consider the following quantity:

$$\begin{aligned}
&\left(v' - \frac{m_s}{M} v_{\chi s}\right)^2 - \left(v - \frac{m_s}{M} v_{\chi s}\right)^2 \\
&= (v' - v)\left[(v' + v) - 2\frac{m_s}{M} v_{\chi s}\right]. \quad (A18)
\end{aligned}$$

Therefore, if $v' \geq v$, we find

$$x_1 = \begin{cases} |v' - \frac{m_s}{M} v_{\chi s}| & \text{if } v_{\chi s} \leq \frac{M}{2m_s}(v + v') \\ |v - \frac{m_s}{M} v_{\chi s}| & \text{otherwise}. \end{cases} \quad (A19)$$

Using $v \leq v'$, the absolute values can be computed explicitly, and we have

$$x_1 = \begin{cases} v' - \frac{m_s}{M} v_{\chi s} & \text{if } v_{\chi s} \leq \frac{M}{2m_s}(v + v') \\ \frac{m_s}{M} v_{\chi s} - v & \text{otherwise}. \end{cases} \quad (A20)$$

With this, we may make the integral in Eq. (A17) fully explicit:

$$\begin{aligned}
\gamma(v, v') &\stackrel{v \leq v'}{=} \int_{\frac{v'+v}{2m_s/M}}^{\infty} dv_{\chi s}\, v_{\chi s}\, \sigma_{\chi s}(v_{\chi s})\, \exp\left[-\frac{m_s m_\chi}{2MT_b} v_{\chi s}^2\right] \\
&\quad \times \left(\mathrm{erf}\left[\sqrt{\frac{M}{2T_b}}\left(\frac{m_s}{M} v_{\chi s} + v\right)\right]\right. \\
&\qquad \left. - \mathrm{erf}\left[\sqrt{\frac{M}{2T_b}}\left(\frac{m_s}{M} v_{\chi s} - v\right)\right]\right) \\
&\quad + \int_{\frac{v'-v}{2m_s/M}}^{\frac{v'+v}{2m_s/M}} dv_{\chi s}\, v_{\chi s}\, \sigma_{\chi s}(v_{\chi s})\, \exp\left[-\frac{m_s m_\chi}{2MT_b} v_{\chi s}^2\right] \\
&\quad \times \left(\mathrm{erf}\left[\sqrt{\frac{M}{2T_b}}\left(v + \frac{m_s}{M} v_{\chi s}\right)\right]\right. \\
&\qquad \left. - \mathrm{erf}\left[\sqrt{\frac{M}{2T_b}}\left(v' - \frac{m_s}{M} v_{\chi s}\right)\right]\right) \quad (A21)
\end{aligned}$$

In summary, for the case where the differential scattering cross-section is isotropic, we have simplified the differential scattering rate to a one-dimensional integral.

### Appendix B: Analytical expressions for $\gamma(v, v')$ for power-law cross-sections with $n \in \{0, 2, 4\}$

We now provide fully analytical expressions for $\gamma(v, v')$ in the case where $\sigma_{\chi s}(v) = \sigma_n v^n$, with $n \in \{0, 2, 4\}$.

First, we rewrite Eq. (A21) in terms of the rescaled velocity $w = \sqrt{M/T_b}\, v$. We moreover change integration variables to $x \equiv (m_s/M)\sqrt{M/T_b}\, v_{\chi s} = m_s v_{\chi s}/\sqrt{MT_b}$. We then obtain

$$\gamma_n(v, v') = \sigma_n \left(\frac{\sqrt{MT_b}}{m_s}\right)^{n+2} \alpha_n(w, w', m_\chi/m_s), \quad (B1)$$

where $\alpha_n(w, w', r)$ is symmetric in $w, w'$, and

$$\begin{aligned}
\alpha_n(w, w', r) &\stackrel{w \leq w'}{=} \int_{\frac{w'+w}{2}}^{\infty} dx\, x^{n+1} e^{-rx^2/2} \\
&\quad \times \left[\mathrm{erf}\left(\frac{x+w}{\sqrt{2}}\right) - \mathrm{erf}\left(\frac{x-w}{\sqrt{2}}\right)\right] \\
&\quad + \int_{\frac{w'-w}{2}}^{\frac{w'+w}{2}} dx\, x^{n+1} e^{-rx^2/2} \\
&\quad \times \left[\mathrm{erf}\left(\frac{x+w}{\sqrt{2}}\right) - \mathrm{erf}\left(\frac{w'-x}{\sqrt{2}}\right)\right]. \quad (B2)
\end{aligned}$$

The functions $\alpha_n$ satisfy the simple recurrence relation:

$$\alpha_{n+2}(w, w', r) = -2\, \partial_r \alpha_n(w, w', r). \quad (B3)$$

We thus start by computing $\alpha_0$, which we may express compactly as follows:

$$\begin{aligned}
\alpha_0(w, w', r) &\stackrel{w \leq w'}{=} \frac{1}{r\sqrt{1+r}} [g(w, w', r) - g(-w, w', r) \\
&\quad + g(w', w, r) - g(w', -w, r)], \quad (B4)
\end{aligned}$$

where
$$g(w_1, w_2, r) \equiv \exp\left[-\frac{r}{r+1}\frac{w_1^2}{2}\right]$$
$$\times \operatorname{erf}\left[\frac{(r-1)w_1 + (r+1)w_2}{2\sqrt{2}\sqrt{r+1}}\right]. \quad (B5)$$

From Eq. (B3), it is then straightforward to obtain analytic expressions for $\alpha_2$ and $\alpha_4$.

Putting everything together, we find that the differential scattering rate can be written as Eq. (34), with

$$\widetilde{\Gamma}(u \to u') = \frac{1}{8c_n}\left(\frac{M^2}{m_s m_\chi}\right)^2 \left(\frac{m_\chi}{m_s}\right)^{n/2+1} \frac{u'}{u} e^{u^2/2}$$
$$\times \alpha_n\left(\sqrt{\frac{M}{m_\chi}}u, \sqrt{\frac{M}{m_\chi}}u', m_\chi/m_s\right). \quad (B6)$$

### Appendix C: 3-D and 1-D velocity diffusion efficiency

In this appendix we compute the average $(\Delta \vec{v})^2$ and $\Delta(v^2)$ per scattering, relevant to our discussion of Eq.s (59) and (60). During an individual scattering with a baryon with velocity $\vec{v}_s$, the DM velocity changes by

$$\vec{v}' = \vec{v} + \frac{m_s}{M}|\vec{v} - \vec{v}_s|(\hat{n}' - \hat{n}), \quad (C1)$$

where $\hat{n}$ and $\hat{n}'$ are the initial and final directions of the DM-baryon relative velocity. In what follows we denote by $\langle X \rangle_Y$ the average of $X$ over the distribution of $Y$.

Assuming for simplicity that the scattering is forward-backward symmetric (the result can be easily generalized otherwise), so that $\langle \hat{n}' \cdot \hat{n} \rangle_{\hat{n}'} = 0$, we then get

$$\langle (\vec{v}' - \vec{v})^2 \rangle_{\hat{n}'} = 2\left(\frac{m_s}{M}\right)^2 |\vec{v} - \vec{v}_s|^2 \quad (C2)$$

Assuming the cross-section scales as $\sigma_{\chi s}(v_{\chi s}) = \sigma_n v_{\chi s}^n$, the average of this quantity per scattering is then

$$\langle (\vec{v}' - \vec{v})^2 \rangle = 2\left(\frac{m_s}{M}\right)^2 \frac{\langle |\vec{v} - \vec{v}_s|^{n+3} \rangle_{\vec{v}_s}}{\langle |\vec{v} - \vec{v}_s|^{n+1} \rangle_{\vec{v}_s}} \quad (C3)$$

For an isotropic MB distribution of baryon velocities, we find

$$\langle |\vec{v} - \vec{v}_s|^p \rangle_{\vec{v}_s} = \frac{1}{2}\frac{1}{(p+2)}\left\langle \frac{(v+v_s)^{p+2} - |v-v_s|^{p+2}}{vv_s}\right\rangle_{v_s}$$
$$= \left(\frac{T_b}{m_s}\right)^{p/2} \lambda_p\left(\sqrt{m_s/T_b}\,v\right), \quad (C4)$$

$$\lambda_p(w) \equiv \frac{1}{p+2}\frac{1}{w}\frac{1}{\sqrt{2\pi}}\int_0^\infty dx\, x\, e^{-x^2/2}$$
$$\times ((x+w)^{p+2} - |x-w|^{p+2})$$
$$= \frac{1}{w}\frac{1}{\sqrt{2\pi}}\int_0^\infty dx\, e^{-x^2/2}$$
$$\times ((x+w)^{p+1} - (x-w)|x-w|^p), \quad (C5)$$

where the second equality was obtained by integrating by parts. Explicity, we have

$$\lambda_p(w) = \frac{1}{w}\frac{1}{\sqrt{2\pi}}\int_0^w dx\, e^{-x^2/2}\left((x+w)^{p+1} + (w-x)^{p+1}\right)$$
$$+ \frac{1}{w}\frac{1}{\sqrt{2\pi}}\int_w^\infty dx\, e^{-x^2/2}\left((x+w)^{p+1} - (x-w)^{p+1}\right). \quad (C6)$$

We therefore have

$$\Delta_{3D}^2(v) \equiv \frac{\langle (\vec{v}' - \vec{v})^2 \rangle}{v^2}$$
$$= 2\left(\frac{m_s}{M}\right)^2 \frac{\lambda_{n+3}(\sqrt{m_s/T_b}\,v)}{(m_s/T_b)v^2\lambda_{n+1}(\sqrt{m_s/T_b}\,v)}. \quad (C7)$$

Let us now evaluate this quantity for a typical initial DM velocity around decoupling, $v \approx \sqrt{T_b/m_\chi}$:

$$\Delta_{3D}^2\left(\sqrt{T_b/m_\chi}\right) = 2\frac{m_s m_\chi}{M^2}\frac{\lambda_{n+3}(\sqrt{m_s/m_\chi})}{\lambda_{n+1}(\sqrt{m_s/m_\chi})}. \quad (C8)$$

Let us now compute the change in velocity *magnitude* squared. Assuming again a forward-backward symmetric scattering, we have

$$\langle v'^2 - v^2 \rangle_{\hat{n}'} = 2\left(\frac{m_s}{M}\right)^2 |\vec{v} - \vec{v}_s|^2 - 2\frac{m_s}{M}(\vec{v} - \vec{v}_s) \cdot \vec{v} \quad (C9)$$

We rewrite the second term as

$$(\vec{v} - \vec{v}_s) \cdot \vec{v} = \frac{1}{2}|\vec{v} - \vec{v}_s|^2 + \frac{1}{2}(v^2 - v_s^2), \quad (C10)$$

so that

$$\langle v'^2 - v^2 \rangle = \frac{m_s(m_s - m_\chi)}{M^2}\frac{\langle |\vec{v} - \vec{v}_s|^{n+3} \rangle_{\vec{v}_s}}{\langle |\vec{v} - \vec{v}_s|^{n+1} \rangle_{\vec{v}_s}} + \frac{m_s}{M}v^2 \Upsilon_n(v),$$
$$\Upsilon_n(v) \equiv \frac{\langle (v_s^2 - v^2)|\vec{v} - \vec{v}_s|^{n+1} \rangle_{\vec{v}_s}}{v^2 \langle |\vec{v} - \vec{v}_s|^{n+1} \rangle_{\vec{v}_s}}. \quad (C11)$$

Computing the average over baryons velocities, we obtain

$$\Upsilon_n(v) = \frac{\kappa_{n+1}(\sqrt{m_s/T_b}\,v)}{(m_s/T_b)v^2\,\lambda_{n+1}(\sqrt{m_s/T_b}\,v)}, \quad (C12)$$

$$\kappa_p(w) \equiv \frac{1}{p+2}\frac{1}{w}\frac{1}{\sqrt{2\pi}}\int_0^\infty dx\, x\, e^{-x^2/2}(x^2 - w^2)$$
$$\times \left((x+w)^{p+2} - |x-w|^{p+2}\right) \quad (C13)$$

From this we may obtain the characteristic relative change of velocity magnitude squared per scattering:

$$\Delta_{1D}^2(v) \equiv \frac{\langle v'^2 - v^2 \rangle}{v^2}. \quad (C14)$$

Again, we evaluate this as $v = \sqrt{T_b/m_\chi}$, and obtain

$$\Delta_{1D}^2\left(\sqrt{T_b/m_\chi}\right) = \frac{(m_s - m_\chi)m_\chi}{M^2}\frac{\lambda_{n+3}(\sqrt{m_s/m_\chi})}{\lambda_{n+1}(\sqrt{m_s/m_\chi})}$$
$$+ \frac{m_\chi}{M}\frac{\kappa_{n+1}(\sqrt{m_s/m_\chi})}{\lambda_{n+1}(\sqrt{m_s/m_\chi})}. \quad (C15)$$